\documentclass{article}
\usepackage{amsfonts}
\usepackage{amsmath}

\newtheorem{theorem}{Theorem}
\newtheorem{acknowledgement}[theorem]{Acknowledgement}

\input{tcilatex}

\begin{document}

\title{Mathematics of structure-function equations of all orders}
\author{Reginald J. Hill \\
National Oceanic and Atmospheric Administration,\\
Environmental Technology Laboratory,\\
Boulder CO 80305-3328, USA}
\date{\today }
\maketitle

\begin{abstract}
Exact equations are derived that relate velocity structure functions of
arbitrary order with other statistics. \ \textquotedblleft
Exact\textquotedblright\ means that no approximation is used except that the
Navier-Stokes equation and incompressibility condition are assumed to be
accurate. \ The exact equations are used to determine the structure-function
equations of all orders for locally homogeneous but anisotropic turbulence
as well as for the locally isotropic case. \ These equations can be used for
investigating the approach to local homogeneity and to local isotropy as
well as the balance of the equations and identification of scaling ranges.
\end{abstract}

\section{INTRODUCTION}

\qquad Full mathematical exposition on the topic of structure-function
equations is given here. \ A brief summary of results derived here will
appear in the Journal of Fluid Mechanics in the paper ``Equations relating
structure functions of all orders.'' \ The two sections below are
sufficiently similar to that paper so as to guide the reader to the relevant
mathematical details, much of which resides in the appendices herein. \ The
two sections below contain more mathematical detail than does that paper. \
The derivation of the structure function equations of all orders produces
substantial mathematical detail. \ This is true for reduction of the viscous
term and for the term involving the pressure gradient when deriving the
exact equations. \ Applying isotropic formulas for structure functions of
arbitrary order requires the invention of new notation and much use of
combinatorial analysis. \ The divergence and Laplacian operating on
isotropic formulas necessarily appear in the equations; evaluation of which
requires the derivation of many identities. \ Finally, matrix-based
algorithms are invented such that the isotropic formulas for the divergence
and Laplacian of isotropic tensors of any order can be generated by computer.

\qquad There is some difference in notation between the paper and this
document. \ In the paper, a component of a structure function is denoted by $%
D_{\left[ N_{1},N_{2},N_{3}\right] }$, whereas here it is denoted by\ the
more complicated notation $D_{\left[ N:N_{1},N_{2},N_{3}\right] }$. \ The
reason for the more complicated notation here is to avoid ambiguity at
several places in the mathematics. \ In the paper, the components of the
tensor $\left\{ \mathbf{W}_{\left[ N-2P\right] }\left( \mathbf{r}\right) 
\mathbf{\delta }_{\left[ 2P\right] }\right\} $\ are denoted by $\left\{ 
\mathbf{W}_{\left[ N-2P\right] }\left( \mathbf{r}\right) \mathbf{\delta }_{%
\left[ 2P\right] }\right\} _{\left[ N_{1},N_{2},N_{3}\right] }$. \ Here,
there is no symbolic distinction between the tensor $\left\{ \mathbf{W}_{%
\left[ N-2P\right] }\left( \mathbf{r}\right) \mathbf{\delta }_{\left[ 2P%
\right] }\right\} $\ and its components. \ The distinction is implied by the
context.

\section{EXACT TWO-POINT EQUATIONS}

\qquad The Navier-Stokes equation for velocity component $u_{i}(\mathbf{x}%
,t) $ and the incompressibility condition are 
\begin{equation}
\partial _{t}u_{i}(\mathbf{x},t)+u_{n}(\mathbf{x},t)\partial _{x_{n}}u_{i}(%
\mathbf{x},t)=-\partial _{x_{i}}p(\mathbf{x},t)+\nu \partial
_{x_{n}}\partial _{x_{n}}u_{i}(\mathbf{x},t)\text{ , and\ }\partial
_{x_{n}}u_{n}(\mathbf{x},t)=0,  \label{1NSE}
\end{equation}
where $p(\mathbf{x},t)$ is the pressure divided by the density (density is
constant), $\nu $ is kinematic viscosity, and $\partial $ denotes partial
differentiation with respect to its subscript variable. \ Summation is
implied by repeated\ Roman indexes. \ Consider another point $\mathbf{x}%
^{\prime }$ such that $\mathbf{x}^{\prime }$\ and $\mathbf{x}$ are
independent variables. \ For brevity, let $u_{i}=u_{i}(\mathbf{x},t)$, $%
u_{i}^{\prime }=u_{i}(\mathbf{x}^{\prime },t)$, etc. Require that $\mathbf{x}
$ and $\mathbf{x}^{\prime }$ have no relative motion. \ Then $\partial
_{x_{i}}u_{j}^{\prime }=0$, $\partial _{x_{i}^{\prime }}u_{j}=0$, etc., and $%
\partial _{t}$ is performed with both \ held $\mathbf{x}$ and $\mathbf{x}%
^{\prime }$ fixed. \ Subtracting (\ref{1NSE}) at $\mathbf{x}^{\prime }$ from
(\ref{1NSE}) at $\mathbf{x} $ and using the aforementioned properties gives 
\begin{eqnarray}
\partial _{t}v_{i}+u_{n}\partial _{x_{n}}v_{i}+u_{n}^{\prime }\partial
_{x_{n}^{\prime }}v_{i} &=&-P_{i}+\nu \left( \partial _{x_{n}}\partial
_{x_{n}}v_{i}+\partial _{x_{n}^{\prime }}\partial _{x_{n}^{\prime
}}v_{i}\right) ,  \label{1firststep} \\
\text{where }v_{i} &\equiv &u_{i}-u_{i}^{\prime }\text{ , }P_{i}\equiv
\left( \partial _{x_{i}}p-\partial _{x_{i}^{\prime }}p^{\prime }\right) .
\end{eqnarray}
Change independent variables from $\mathbf{x}$ and $\mathbf{x}^{\prime }$ to
the sum and difference independent variables: 
\begin{equation}
\mathbf{X}\equiv \left( \mathbf{x}+\mathbf{x}^{\prime }\right) /2\text{ \
and \ }\mathbf{r}\equiv \mathbf{x}-\mathbf{x}^{\prime },\ \ \text{and define 
}r\equiv \left| \mathbf{r}\right| .
\end{equation}
The relationship between the partial derivatives is 
\begin{equation}
\partial _{x_{i}}=\partial _{r_{i}}+\frac{1}{2}\partial _{X_{i}}\text{ , }\
\partial _{x_{i}^{\prime }}=-\partial _{r_{i}}+\frac{1}{2}\partial _{X_{i}}%
\text{ \ , }\partial _{X_{i}}=\partial _{x_{i}}+\partial _{x_{i}^{\prime }}%
\text{ \ , }\partial _{r_{i}}=\frac{1}{2}\left( \partial _{x_{i}}-\partial
_{x_{i}^{\prime }}\right) .  \label{1identderivs}
\end{equation}
The change of variables organizes the equations in a revealing way because
of the following properties. \ In the case of homogeneous turbulence, $%
\partial _{X_{i}}$ operating on a statistic produces zero because that
derivative is the rate of change with respect to the place where the
measurement is performed. \ Consider a term in an equation composed of $%
\partial _{X_{i}}$ operating on a statistic. \ For locally homogeneous
turbulence, that term becomes negligible as $r$ is decreased relative to the
integral scale. \ For the homogeneous and locally homogeneous cases, the
statistical equations retain their dependence on $\mathbf{r}$, which is the
displacement vector of two points of measurement. \ Using (\ref{1identderivs}%
), (\ref{1firststep}) becomes 
\begin{eqnarray}
\partial _{t}v_{i}+U_{n}\partial _{X_{n}}v_{i}+v_{n}\partial _{r_{n}}v_{i}
&=&-P_{i}+\nu \left( \partial _{x_{n}}\partial _{x_{n}}v_{i}+\partial
_{x_{n}^{\prime }}\partial _{x_{n}^{\prime }}v_{i}\right) ,  \label{2ndstep}
\\
\text{where }U_{n} &\equiv &\left( u_{i}+u_{i}^{\prime }\right) /2.
\end{eqnarray}

\qquad Now multiply (\ref{2ndstep}) by the product $v_{j}v_{k}\cdot \cdot
\cdot v_{l}$, which contains $N-1$ factors of velocity difference, each
factor having a distinct index. \ Sum the $N$ such equations as required to
produce symmetry under interchange of each pair of indexes, excluding the
summation index $n$. \ French braces, i.e., $\left\{ \circ \right\} $,
denote the sum of all terms of a given type that produce symmetry under
interchange of each pair of indexes. \ The differentiation chain rule gives 
\begin{eqnarray}
\left\{ v_{j}v_{k}\cdot \cdot \cdot v_{l}\partial _{t}v_{i}\right\}
&=&\partial _{t}\left( v_{j}v_{k}\cdot \cdot \cdot v_{l}v_{i}\right) , \\
\left\{ v_{j}v_{k}\cdot \cdot \cdot v_{l}U_{n}\partial _{X_{n}}v_{i}\right\}
&=&U_{n}\partial _{X_{n}}\left( v_{j}v_{k}\cdot \cdot \cdot
v_{l}v_{i}\right) =\partial _{X_{n}}\left( U_{n}v_{j}v_{k}\cdot \cdot \cdot
v_{l}v_{i}\right) ,  \label{1p} \\
\left\{ v_{j}v_{k}\cdot \cdot \cdot v_{l}v_{n}\partial _{r_{n}}v_{i}\right\}
&=&v_{n}\partial _{r_{n}}\left( v_{j}v_{k}\cdot \cdot \cdot
v_{l}v_{i}\right) =\partial _{r_{n}}\left( v_{n}v_{j}v_{k}\cdot \cdot \cdot
v_{l}v_{i}\right) .  \label{1pa}
\end{eqnarray}%
The right-most expressions in (\ref{1p}) and (\ref{1pa}) follow from the
incompressibility property obtained from (\ref{1identderivs}) and the fact
that $\partial _{x_{i}}u_{j}^{\prime }=0$, $\partial _{x_{i}^{\prime
}}u_{j}=0$, namely 
\begin{equation}
\partial _{X_{n}}U_{n}=0,\partial _{X_{n}}v_{n}=0,\partial
_{r_{n}}U_{n}=0,\partial _{r_{n}}v_{n}=0.
\end{equation}%
The viscous term in (\ref{2ndstep}) produces $\nu \left\{ v_{j}v_{k}\cdot
\cdot \cdot v_{l}\left( \partial _{x_{n}}\partial _{x_{n}}v_{i}+\partial
_{x_{n}^{\prime }}\partial _{x_{n}^{\prime }}v_{i}\right) \right\} $; this
expression is treated in Appendix A. \ These results give 
\begin{eqnarray}
&&\partial _{t}\left( v_{j}\cdot \cdot \cdot v_{i}\right) +\partial
_{X_{n}}\left( U_{n}v_{j}\cdot \cdot \cdot v_{i}\right) +\partial
_{r_{n}}\left( v_{n}v_{j}\cdot \cdot \cdot v_{i}\right) =  \notag \\
&&-\left\{ v_{j}\cdot \cdot \cdot v_{l}P_{i}\right\} \\
&&+2\nu \left[ \left( \partial _{r_{n}}\partial _{r_{n}}+\frac{1}{4}\partial
_{X_{n}}\partial _{X_{n}}\right) \left( v_{j}\cdot \cdot \cdot v_{i}\right)
-\left\{ v_{k}\cdot \cdot \cdot v_{l}e_{ij}\right\} \right] ,
\label{1exactunave} \\
\text{where }e_{ij} &\equiv &\left( \partial _{x_{n}}u_{i}\right) \left(
\partial _{x_{n}}u_{j}\right) +\left( \partial _{x_{n}^{\prime
}}u_{i}^{\prime }\right) \left( \partial _{x_{n}^{\prime }}u_{j}^{\prime
}\right) =\left( \partial _{x_{n}}v_{i}\right) \left( \partial
_{x_{n}}v_{j}\right) +\left( \partial _{x_{n}^{\prime }}v_{i}\right) \left(
\partial _{x_{n}^{\prime }}v_{j}\right) .  \notag
\end{eqnarray}

\qquad The quantity $\left\{ v_{j}\cdot \cdot \cdot v_{l}P_{i}\right\} $\
can be expressed differently on the basis that (\ref{1identderivs}) allows $%
P_{i}$\ to be written as $P_{i}=\partial _{X_{i}}\left( p-p^{\prime }\right) 
$. The derivation is in Appendix B; the alternative formula is 
\begin{equation}
\left\{ v_{j}v_{k}\cdot \cdot \cdot v_{l}P_{i}\right\} =\left\{ \partial
_{X_{i}}\left[ v_{j}v_{k}\cdot \cdot \cdot v_{l}\left( p-p^{\prime }\right) %
\right] \right\} -2\left( N-1\right) \left( p-p^{\prime }\right) \left\{
\left( s_{ij}-s_{ij}^{\prime }\right) v_{k}\cdot \cdot \cdot v_{l}\right\} ,
\label{1Palternative}
\end{equation}%
where the rate of strain tensor $s_{ij}$\ is defined by 
\begin{equation}
s_{ij}\equiv \left( \partial _{x_{i}}u_{j}+\partial _{x_{j}}u_{i}\right) /2.
\label{1straindefine}
\end{equation}

\section{AVERAGED EQUATIONS}

\qquad Consider the ensemble average because it commutes with temporal and
spatial derivatives. \ The above notation of explicit indexes is burdensome.
\ Because the tensors are symmetric, it suffices to show only the number of
indexes. \ Define the following statistical tensors, which are symmetric
under interchange of any pair of indexes, excluding the summation index $n$
in the definition of $\mathbf{F}_{\left[ N+1\right] }$: 
\begin{equation}
\mathbf{D}_{\left[ N\right] }\equiv \left\langle v_{j}\cdot \cdot \cdot
v_{i}\right\rangle ,\mathbf{F}_{\left[ N+1\right] }\equiv \left\langle
U_{n}v_{j}\cdot \cdot \cdot v_{i}\right\rangle ,\mathbf{T}_{\left[ N\right]
}\equiv \left\langle \left\{ v_{j}\cdot \cdot \cdot v_{l}P_{i}\right\}
\right\rangle ,\mathbf{E}_{\left[ N\right] }\equiv \left\langle \left\{
v_{k}\cdot \cdot \cdot v_{l}e_{ij}\right\} \right\rangle ,
\label{1implicitindex}
\end{equation}
where angle brackets $\left\langle {}\right\rangle $ denote the ensemble
average, and the subscripts $N$\ and $N+1$ within square brackets denote the
number of indexes. \ The argument list $\left( \mathbf{X},\mathbf{r}%
,t\right) $ is understood for each tensor. \ The left-hand sides of each
definition in (\ref{1implicitindex}) are in implicit-index notation for
which only the number of indexes is given; the right-hand sides in (\ref%
{1implicitindex}) are in explicit-index notation. \ The ensemble average of (%
\ref{1exactunave}) is 
\begin{equation}
\partial _{t}\mathbf{D}_{\left[ N\right] }+\nabla _{\mathbf{X}}\cdot \mathbf{%
F}_{\left[ N+1\right] }+\nabla _{\mathbf{r}}\cdot \mathbf{D}_{\left[ N+1%
\right] }=-\mathbf{T}_{\left[ N\right] }+2\nu \left[ \left( \nabla _{\mathbf{%
r}}^{2}+\frac{1}{4}\nabla _{\mathbf{X}}^{2}\right) \mathbf{D}_{\left[ N%
\right] }-\mathbf{E}_{\left[ N\right] }\right] ,  \label{2implicitindex}
\end{equation}
where, $\nabla _{\mathbf{X}}\cdot \mathbf{F}_{\left[ N+1\right] }\equiv
\partial _{X_{n}}\left\langle U_{n}v_{j}\cdot \cdot \cdot v_{i}\right\rangle
,\nabla _{\mathbf{r}}\cdot \mathbf{D}_{\left[ N+1\right] }\equiv \partial
_{r_{n}}\left\langle v_{n}v_{j}\cdot \cdot \cdot v_{i}\right\rangle ,\nabla
_{\mathbf{r}}^{2}\equiv \partial _{r_{n}}\partial _{r_{n}},\nabla _{\mathbf{X%
}}^{2}\equiv \partial _{X_{n}}\partial _{X_{n}}$. \ The notations $\nabla _{%
\mathbf{X}}\cdot $, $\nabla _{\mathbf{X}}^{2}$, $\nabla _{\mathbf{r}}\cdot $%
, and $\nabla _{\mathbf{r}}^{2}$ are the divergence and Laplacian operators
in $\mathbf{X} $-space and $\mathbf{r}$-space, respectively.

\subsection{HOMOGENEOUS AND LOCALLY HOMOGENEOUS TURBULENCE}

\qquad Consider homogeneous turbulence and locally homogeneous turbulence;
the latter applies for small $r$ and large Reynolds number. \ The variation
of the statistics with the location of measurement or of evaluation is
neglected for these cases. \ That location being $\mathbf{X}$\textbf{,} the
result of $\nabla _{\mathbf{X}}\cdot $ operating on a statistic is
neglected. \ Thus the terms $\nabla _{\mathbf{X}}\cdot \mathbf{F}_{\left[ N+1%
\right] }$\ and $\frac{1}{4}\nabla _{\mathbf{X}}^{2}\mathbf{D}_{\left[ N%
\right] }$ are neglected in (\ref{2implicitindex}); then (\ref%
{2implicitindex}) becomes 
\begin{equation}
\partial _{t}\mathbf{D}_{\left[ N\right] }+\nabla _{\mathbf{r}}\cdot \mathbf{%
D}_{\left[ N+1\right] }=-\mathbf{T}_{\left[ N\right] }+2\nu \left[ \nabla _{%
\mathbf{r}}^{2}\mathbf{D}_{\left[ N\right] }-\mathbf{E}_{\left[ N\right] }%
\right] .  \label{1iso}
\end{equation}%
Because the $\mathbf{X}$-dependence is neglected, the argument list $\left( 
\mathbf{r},t\right) $ is understood for each tensor. \ The ensemble average
of (\ref{1Palternative}) contains $\left\langle \partial _{X_{i}}\left[
\left\{ v_{j}v_{k}\cdot \cdot \cdot v_{l}\left( p-p^{\prime }\right)
\right\} \right] \right\rangle $, which can be written as the sum of $N-1$
statistics of the form $\left\langle \left\{ v_{j}v_{k}\cdot \cdot \cdot
v_{l}\left( p-p^{\prime }\right) \right\} \right\rangle $\ operated upon by
the $\mathbf{X}$-space\ gradient. \ Since such $\mathbf{X}$-space derivative
terms are neglected, (\ref{1Palternative}) gives the alternative that 
\begin{equation}
\mathbf{T}_{\left[ N\right] }=-2\left( N-1\right) \left\langle \left(
p-p^{\prime }\right) \left\{ \left( s_{ij}-s_{ij}^{\prime }\right)
v_{k}\cdot \cdot \cdot v_{l}\right\} \right\rangle .
\end{equation}%
Locally homogeneous turbulence is also locally stationary such that the term 
$\partial _{t}\mathbf{D}_{\left[ N\right] }$\ in (\ref{1iso}) may be
neglected. \ However, $\partial _{t}\mathbf{D}_{\left[ N\right] }$\ is not
necessarily negligible for homogeneous turbulence.

\subsection{ISOTROPIC AND LOCALLY ISOTROPIC\ TURBULENCE}

\qquad Consider isotropic turbulence and locally isotropic turbulence; the
latter applies for small $r$ and large Reynolds number. \ The tensors $%
\mathbf{D}_{\left[ N\right] }$, $\mathbf{T}_{\left[ N\right] }$, and $%
\mathbf{E}_{\left[ N\right] }$\ in (\ref{1implicitindex}) obey the isotropic
formula. \ The Kronecker delta $\delta _{ij}=1$ if $i=j$ and $\delta _{ij}=0$
if $i\neq j$. $\ $\ Let $\mathbf{\delta }_{\left[ 2P\right] }$ denote the
product of $P$ Kronecker deltas having $2P$ distinct indexes, and let $%
\mathbf{W}_{\left[ N\right] }\left( \mathbf{r}\right) $ denote the product
of $N$ factors $\frac{r_{i}}{r}$ each with a distinct index; the argument $%
\mathbf{r}$\ is omitted when clarity does not suffer. \ Because each tensor
in (\ref{1implicitindex}) is symmetric under interchange of any two indexes,
their isotropic formulas are particularly simple. \ Each formula is a the
sum of $M+1$\ terms where 
\begin{equation}
M=N/2\text{ if }N\text{ is even, and }M=\left( N-1\right) /2\text{ if }N%
\text{ is odd.}  \label{1M}
\end{equation}
Each term is the product of a distinct scalar function with a $\mathbf{W}_{%
\left[ N\right] }$\ and a $\mathbf{\delta }_{\left[ 2P\right] }$. \ From one
term to the next, a pair of indexes is transferred from a $\mathbf{W}_{\left[
N\right] }$ to a $\mathbf{\delta }_{\left[ 2P\right] }$; examples are given
in (\ref{isoexample}-\ref{isoexample3}) of Appendix E. \ For the tensor $%
\mathbf{D}_{\left[ N\right] }$, denote the $P$th scalar function by $%
D_{N,P}\left( r,t\right) $. \ Thus the scalar functions belonging to the
isotropic formulas for $\mathbf{T}_{\left[ N\right] }$, $\mathbf{E}_{\left[ N%
\right] }$, and $\mathbf{D}_{\left[ N+1\right] }$ are denoted by $%
T_{N,P}\left( r,t\right) $, $E_{N,P}\left( r,t\right) $, and $%
D_{N+1,P}\left( r,t\right) $, respectively. \ The scalar functions depend on
the magnitude of the spacing $r$ rather than on the vector spacing $\mathbf{r%
}$. \ The isotropic formula for $\mathbf{D}_{\left[ N\right] }$\ is 
\begin{equation}
\mathbf{D}_{\left[ N\right] }\left( \mathbf{r},t\right) =\overset{M}{%
\underset{P=0}{\sum }}D_{N,P}\left( r,t\right) \left\{ \mathbf{W}_{\left[
N-2P\right] }\left( \mathbf{r}\right) \mathbf{\delta }_{\left[ 2P\right]
}\right\} ,  \label{1isoformula}
\end{equation}
and the isotropic formulas for $\mathbf{T}_{\left[ N\right] }$ and $\mathbf{E%
}_{\left[ N\right] }$ have the analogous notation. \ Recall that $\left\{
\circ \right\} $ denotes the sum of all terms of a given type that produce
symmetry under interchange of each pair of indexes. \ Henceforth, the
argument list $\left( r,t\right) $\ will be deleted.

\qquad A special Cartesian coordinate system is typically used because it
simplifies the isotropic formula. \ This coordinate system has the positive $%
1$-axis parallel to the direction of $\mathbf{r}$, and the $2$- and $3$-axes
are therefore perpendicular to $\mathbf{r}$. \ Let $N_{1}$, $N_{2}$, and $%
N_{3}$ be the number of indexes of a component of $\mathbf{D}_{\left[ N%
\right] }$ that are $1$, $2$, and $3$, respectively; such that $%
N=N_{1}+N_{2}+N_{3}$. \ Because of symmetry, the order of indexes is
immaterial such that a component of $\mathbf{D}_{\left[ N\right] }$ can be
identified by $N_{1}$, $N_{2}$, and $N_{3}$. \ Thus, denote a component of $%
\mathbf{D}_{\left[ N\right] }$ by $D_{\left[ N:N_{1},N_{2},N_{3}\right] }$,
which is a function of $\mathbf{r}$ and $t$. \ The projection of (\ref%
{1isoformula}) using $N_{1}$, $N_{2}$, and $N_{3}$\ unit vectors in the
directions of the $1$-, $2$-, and $3$-axes, respectively, results in the
component $D_{\left[ N:N_{1},N_{2},N_{3}\right] }$\ on the left-hand side of
(\ref{1isoformula}), and numerical values of the projection of $\left\{ 
\mathbf{W}_{\left[ N-2P\right] }\left( \mathbf{r}\right) \mathbf{\delta }_{%
\left[ 2P\right] }\right\} $\ appear on the right-hand side. \ Henceforth
the word ''projection'' will be omitted for brevity. Those values of the
coefficients $\left\{ \mathbf{W}_{\left[ N-2P\right] }\left( \mathbf{r}%
\right) \mathbf{\delta }_{\left[ 2P\right] }\right\} $\ in (\ref{1isoformula}%
) are needed; the values obtained for the special coordinate system are
determined in Appendix C; they are, from (\ref{AppC1}-\ref{AppC2}),

\begin{eqnarray}
\text{if }2P &<&N_{2}+N_{3}\text{ then }\left\{ \mathbf{W}_{\left[ N-2P%
\right] }\mathbf{\delta }_{\left[ 2P\right] }\right\} =0\text{; otherwise,}
\label{0sinisocoefs} \\
&&\left\{ \mathbf{W}_{\left[ N-2P\right] }\mathbf{\delta }_{\left[ 2P\right]
}\right\}  \notag \\
&=&N_{1}!N_{2}!N_{3}!/\left[ \left( N-2P\right) !2^{P}\left( \frac{N_{2}}{2}%
\right) !\left( \frac{N_{3}}{2}\right) !\left( P-\frac{N_{2}}{2}-\frac{N_{3}%
}{2}\right) !\right] .  \label{1isocoefs}
\end{eqnarray}

\qquad By applying (\ref{1isoformula}) and (\ref{0sinisocoefs}-\ref%
{1isocoefs}) for all combinations of indexes, one can determine which
components $D_{\left[ N:N_{1},N_{2},N_{3}\right] }$\ are zero and which are
nonzero, identify $M+1$\ linearly independent equations that determine\ the $%
D_{N,P}$ in terms of $M+1$ of the $D_{\left[ N:N_{1},N_{2},N_{3}\right] }$,\
and find algebraic relationships between the remaining nonzero $D_{\left[
N:N_{1},N_{2},N_{3}\right] }$. \ The derivations are in Appendix D; a
summary follows.

\qquad A component $D_{\left[ N:N_{1},N_{2},N_{3}\right] }$ is nonzero only
if both $N_{2}$ and $N_{3}$ are even and when $N_{1}$ is odd if $N$ is odd
and when $N_{1}$ is even if $N$ is even. \ Thereby, $\left( M+1\right)
\left( M+2\right) /2$\ components are nonzero. \ There are $3^{N}$
components of $\mathbf{D}_{\left[ N\right] }$; thus the other $3^{N}-\left(
M+1\right) \left( M+2\right) /2$ components are zero.

\qquad There exists exactly $\left( M+1\right) M/2$\ kinematic relationships
among the nonzero components of $\mathbf{D}_{\left[ N\right] }$. \ For each
of the $M+1$ cases of $N_{1}$, these relationships are expressed by the
proportionality 
\begin{eqnarray}
D_{\left[ N:N_{1},2L,0\right] } &:&D_{\left[ N:N_{1},2L-2,2\right] }:D_{%
\left[ N:N_{1},2L-4,4\right] }:\cdot \cdot \cdot :D_{\left[ N:N_{1},0,2L%
\right] }=  \notag \\
\left[ \left( 2L\right) !0!/L!0!\right] &:&\left[ \left( 2L-2\right)
!2!/\left( L-1\right) !1!\right] :\left[ \left( 2L-4\right) !4!/\left(
L-2\right) !2!\right] : \\
&:&\cdot \cdot \cdot :\left[ 0!\left( 2L\right) !/0!L!\right] .
\label{Kinematic}
\end{eqnarray}%
Previously, only one such kinematic relationship was known (Millionshtchikov
1941). \ For $N=4$, (\ref{Kinematic}) gives $D_{\left[ 4:0,4,0\right] }:D_{%
\left[ 4:0,2,2\right] }:D_{\left[ 4:0,0,4\right] }=12:4:12$. \ In
explicit-index notation this can be written as $D_{2222}=3D_{2233}=D_{3333}$%
, which was discovered by Millionshtchikov (1941). \ Now, all such
relationships are known.

\qquad There remain $M+1$\ linearly independent nonzero components of $%
\mathbf{D}_{\left[ N\right] }$. \ This must be so because there are $M+1$\
terms\ in (\ref{1isoformula}), and the $M+1$ scalar functions $D_{N,P}$\
therein must be related to $M+1$\ components. \ Consider the $M+1$\ linearly
independent equations that determine\ the $D_{N,P}$ in terms of $M+1$ of the 
$D_{\left[ N:N_{1},N_{2},N_{3}\right] }$. \ For simplicity, the chosen
components can all have $N_{3}=0$, i.e., the choice of linearly independent
components can be $D_{\left[ N:N,0,0\right] }$, $D_{\left[ N:N-2,2,0\right]
} $, $D_{\left[ N:N-4,4,0\right] }$, $\cdot \cdot \cdot $, $D_{\left[
N:N-2M,2M,0\right] }$. \ As described above, projections of (\ref%
{1isoformula}) result in the chosen components on the left-hand side and
algebraic equations on the right-hand side. \ These equations can be
expressed in matrix form and solved by matrix inversion methods; the result
is given in (\ref{1Dsolveq}) of Appendix F. \ Given experimental or DNS data
or a theoretical formula for the chosen components, the solution in (\ref%
{1Dsolveq}) determines the functions $D_{N,P}$\ in (\ref{1isoformula}); then
(\ref{1isoformula}) completely specifies the tensor $\mathbf{D}_{\left[ N%
\right] }$. \ The matrix algorithm is an efficient means of determining
isotropic expressions for the terms $\nabla _{\mathbf{r}}\cdot \mathbf{D}_{%
\left[ N+1\right] }$\ and $\nabla _{\mathbf{r}}^{2}\mathbf{D}_{\left[ N%
\right] }$\ in (\ref{1iso}). \ Those algorithms are given in Appendix F. \
From the example for $N=2$ in Appendix F, use of the matrix algorithm and
the isotropic formulas in (\ref{1iso}) gives the two scalar equations 
\begin{eqnarray}
&&\partial _{t}D_{11}+\left( \partial _{r}+\frac{2}{r}\right) D_{111}-\frac{4%
}{r}D_{122}  \notag \\
&=&-T_{11}+2\nu \left[ \left( \partial _{r}^{2}+\frac{2}{r}\partial _{r}-%
\frac{4}{r^{2}}\right) D_{11}+\frac{4}{r^{2}}D_{22}-E_{11}\right]  \notag \\
&=&2\nu \left[ \partial _{r}^{2}D_{11}+\frac{2}{r}\partial _{r}D_{11}+\frac{4%
}{r^{2}}\left( D_{22}-D_{11}\right) \right] -4\varepsilon /3,
\label{2ndorder} \\
&&\partial _{t}D_{22}+\left( \partial _{r}+\frac{4}{r}\right) D_{122}  \notag
\\
&=&-T_{22}+2\nu \left[ \frac{2}{r^{2}}D_{11}+\left( \partial _{r}^{2}+\frac{2%
}{r}\partial _{r}-\frac{2}{r^{2}}\right) D_{22}-E_{22}\right]  \notag \\
&=&2\nu \left[ \partial _{r}^{2}D_{22}+\frac{2}{r}\partial _{r}D_{22}-\frac{2%
}{r^{2}}\left( D_{22}-D_{11}\right) \right] -4\varepsilon /3,
\label{2ndorder2}
\end{eqnarray}%
where use was made of the fact (Hill, 1997) that local isotropy gives $%
T_{11}=T_{22}=0$ and $2\nu E_{11}=2\nu E_{22}=4\varepsilon /3$ where $%
\varepsilon $\ is the average energy dissipation rate per unit mass of
fluid. \ Now, (\ref{2ndorder}-\ref{2ndorder2}) are the same as equations
(43-44) of Hill (1997), and Hill (1997) shows how these equations lead to
Kolmogorov's equation and his 4/5 law. \ From the example for $N=3$ in
Appendix F, 
\begin{eqnarray}
\partial _{t}D_{111}+\left( \partial _{r}+\frac{2}{r}\right) D_{1111}-\frac{6%
}{r}D_{1122} &=&-T_{111}+2\nu \left[ C-E_{111}\right] ,  \label{HillBor1} \\
\partial _{t}D_{122}+\left( \partial _{r}+\frac{4}{r}\right) D_{1122}-\frac{4%
}{3r}D_{2222} &=&-T_{122}+2\nu \left[ B-E_{122}\right] ,  \label{HillBor2}
\end{eqnarray}%
where 
\begin{equation}
C\equiv \left( -\frac{4}{r^{2}}+\frac{4}{r}\partial _{r}+\partial
_{r}^{2}\right) D_{111},\text{ and }B\equiv \frac{1}{6}\left( \frac{4}{r^{2}}%
-\frac{4}{r}\partial _{r}+5\partial _{r}^{2}+r\partial _{r}^{3}\right)
D_{111}.  \label{HillBor3}
\end{equation}%
The incompressibility condition, $D_{122}=\frac{1}{6}\left(
D_{111}+r\partial _{r}D_{111}\right) $, was substituted in (\ref{LapNis3})
to obtain (\ref{HillBor3}). \ The matrix algorithm is checked by the fact
that (\ref{HillBor1}-\ref{HillBor3}) are the same as given by Hill and
Boratav (2001).

\qquad The equations for $N=4$\ are

\begin{equation*}
\partial _{t}D_{1111}+\left( \partial _{r}+\frac{2}{r}\right) D_{11\,111}-%
\frac{8}{r}D_{11\,122}=
\end{equation*}%
\begin{equation}
=-T_{1111}+2\nu \left[ \left( \partial _{r}^{2}+\frac{2}{r}\partial _{r}-%
\frac{8}{r^{2}}\right) D_{1111}+\frac{24}{r^{2}}D_{1122}\right] -2\nu
E_{1111},
\end{equation}%
\begin{equation*}
\partial _{t}D_{1122}+\left( \partial _{r}+\frac{4}{r}\right) D_{11\,122}-%
\frac{8}{3r}D_{12\,222}=
\end{equation*}%
\begin{equation}
=-T_{1122}+2\nu \left[ \frac{2}{r^{2}}D_{1111}+\left( \partial _{r}^{2}+%
\frac{2}{r}\partial _{r}-\frac{14}{r^{2}}\right) D_{1122}+\frac{8}{3r^{2}}%
D_{2222}\right] -2\nu E_{1122},
\end{equation}%
\begin{equation}
\partial _{t}D_{2222}+\left( \partial _{r}+\frac{6}{r}\right)
D_{12\,222}=-T_{2222}+2\nu \left[ \frac{12}{r^{2}}D_{1122}+\left( \partial
_{r}^{2}+\frac{2}{r}\partial _{r}-\frac{4}{r^{2}}\right) D_{2222}\right]
-2\nu E_{2222}.
\end{equation}

Since these equations have a repetitive structure, it suffices to give the
divergence and Laplacian terms. \ Simplify index notation such that
subscript $\left[ N_{1},N_{2},N_{3}\right] $ denotes $N_{1}$ subscripts$1$
and $N_{2}$\ subscripts $2$ and $N_{3}$\ subscripts $3$. \ For $N=4$\ to $8$%
\ the divergence and Laplacian operators are, respectively:

\begin{equation*}
N=4
\end{equation*}%
\bigskip $\left( 
\begin{array}{c}
\left( \partial _{r}+\frac{2}{r}\right) D_{\left[ 5,0,0\right] }-\frac{8}{r}%
D_{\left[ 3,2,0\right] } \\ 
\left( \partial _{r}+\frac{4}{r}\right) D_{\left[ 3,2,0\right] }-\frac{8}{3r}%
D_{\left[ 1,4,0\right] } \\ 
\left( \partial _{r}+\frac{6}{r}\right) D_{\left[ 1,4,0\right] }%
\end{array}%
\right) $ $\ \ \ \ \left( \ 
\begin{array}{c}
\left( \partial _{r}^{2}+\frac{2}{r}\partial _{r}-\frac{8}{r^{2}}\right) D_{%
\left[ 4,0,0\right] }+\frac{24}{r^{2}}D_{\left[ 2,2,0\right] } \\ 
\left( \partial _{r}^{2}+\frac{2}{r}\partial _{r}-\frac{14}{r^{2}}\right) D_{%
\left[ 2,2,0\right] }+\frac{2}{r^{2}}D_{\left[ 4,0,0\right] }+\frac{8}{3r^{2}%
}D_{\left[ 0,4,0\right] } \\ 
\left( \partial _{r}^{2}+\frac{2}{r}\partial _{r}-\frac{4}{r^{2}}\right) D_{%
\left[ 0,4,0\right] }+\frac{12}{r^{2}}D_{\left[ 2,2,0\right] }%
\end{array}%
\right) $

------------------------------------------------------------------------------------------------------%
\begin{equation*}
N=5
\end{equation*}%
$\left( 
\begin{array}{c}
\left( \partial _{r}+\frac{2}{r}\right) D_{\left[ 6,0,0\right] }-\frac{10}{r}%
D_{\left[ 4,2,0\right] } \\ 
\left( \partial _{r}+\frac{4}{r}\right) D_{\left[ 4,2,0\right] }-\frac{4}{r}%
D_{\left[ 2,4,0\right] } \\ 
\left( \partial _{r}+\frac{6}{r}\right) D_{\left[ 2,4,0\right] }-\frac{6}{5r}%
D_{\left[ 0,6,0\right] }%
\end{array}%
\right) $ \ \ \ \ \ $\left( \ 
\begin{array}{c}
\left( \partial _{r}^{2}+\frac{2}{r}\partial _{r}-\frac{10}{r^{2}}\right) D_{%
\left[ 5,0,0\right] }+\frac{40}{r^{2}}D_{\left[ 3,2,0\right] } \\ 
\left( \partial _{r}^{2}+\frac{2}{r}\partial _{r}-\frac{20}{r^{2}}\right) D_{%
\left[ 3,2,0\right] }+\frac{2}{r^{2}}D_{\left[ 5,0,0\right] }+\frac{8}{r^{2}}%
D_{\left[ 1,4,0\right] } \\ 
\left( \partial _{r}^{2}+\frac{2}{r}\partial _{r}-\frac{14}{r^{2}}\right) D_{%
\left[ 1,4,0\right] }+\frac{12}{r^{2}}D_{\left[ 3,2,0\right] }%
\end{array}%
\right) $

------------------------------------------------------------------------------------------------------%
\begin{equation*}
N=6
\end{equation*}%
$\left( 
\begin{array}{c}
\left( \partial _{r}+\frac{2}{r}\right) D_{\left[ 7,0,0\right] }-\frac{12}{r}%
D_{\left[ 5,2,0\right] } \\ 
\left( \partial _{r}+\frac{4}{r}\right) D_{\left[ 5,2,0\right] }-\frac{16}{3r%
}D_{\left[ 3,4,0\right] } \\ 
\left( \partial _{r}+\frac{6}{r}\right) D_{\left[ 3,4,0\right] }-\frac{12}{5r%
}D_{\left[ 1,6,0\right] } \\ 
\left( \partial _{r}+\frac{8}{r}\right) D_{\left[ 1,6,0\right] }%
\end{array}%
\right) $ \ \ \ \ \ $\left( 
\begin{array}{c}
\left( \partial _{r}^{2}+\frac{2}{r}\partial _{r}-\frac{12}{r^{2}}\right) D_{%
\left[ 6,0,0\right] }+\frac{60}{r^{2}}D_{\left[ 4,2,0\right] } \\ 
\left( \partial _{r}^{2}+\frac{2}{r}\partial _{r}-\frac{26}{r^{2}}\right) D_{%
\left[ 4,2,0\right] }+\frac{2}{r^{2}}D_{\left[ 6,0,0\right] }+\frac{16}{r^{2}%
}D_{\left[ 2,4,0\right] } \\ 
\left( \partial _{r}^{2}+\frac{2}{r}\partial _{r}-\frac{24}{r^{2}}\right) D_{%
\left[ 2,4,0\right] }+\frac{12}{r^{2}}D_{\left[ 4,2,0\right] }+\frac{12}{%
5r^{2}}D_{\left[ 0,6,0\right] } \\ 
\left( \partial _{r}^{2}+\frac{2}{r}\partial _{r}-\frac{6}{r^{2}}\right) D_{%
\left[ 0,6,0\right] }+\frac{30}{r^{2}}D_{\left[ 2,4,0\right] }%
\end{array}%
\right) $

------------------------------------------------------------------------------------------------------%
\begin{equation*}
N=7
\end{equation*}

$\left( \bigskip 
\begin{array}{c}
\left( \partial _{r}+\frac{2}{r}\right) D_{\left[ 8,0,0\right] }-\frac{14}{r}%
D_{\left[ 6,2,0\right] } \\ 
\left( \partial _{r}+\frac{4}{r}\right) D_{\left[ 6,2,0\right] }-\frac{20}{3r%
}D_{\left[ 4,4,0\right] } \\ 
\left( \partial _{r}+\frac{6}{r}\right) D_{\left[ 4,4,0\right] }-\frac{18}{5r%
}D_{\left[ 2,6,0\right] } \\ 
\left( \partial _{r}+\frac{8}{r}\right) D_{\left[ 2,6,0\right] }-\frac{8}{7r}%
D_{\left[ 0,8,0\right] }%
\end{array}%
\right) $\ \ $\left( 
\begin{array}{c}
\left( \partial _{r}^{2}+\frac{2}{r}\partial _{r}-\frac{14}{r^{2}}\right) D_{%
\left[ 7,0,0\right] }+\frac{84}{r^{2}}D_{\left[ 5,2,0\right] } \\ 
\left( \partial _{r}^{2}+\frac{2}{r}\partial _{r}-\frac{32}{r^{2}}\right) D_{%
\left[ 5,2,0\right] }+\frac{2}{r^{2}}D_{\left[ 7,0,0\right] }+\frac{80}{%
3r^{2}}D_{\left[ 3,4,0\right] } \\ 
\left( \partial _{r}^{2}+\frac{2}{r}\partial _{r}-\frac{34}{r^{2}}\right) D_{%
\left[ 3,4,0\right] }+\frac{12}{r^{2}}D_{\left[ 5,2,0\right] }+\frac{36}{%
5r^{2}}D_{\left[ 1,6,0\right] } \\ 
\left( \partial _{r}^{2}+\frac{2}{r}\partial _{r}-\frac{20}{r^{2}}\right) D_{%
\left[ 1,6,0\right] }+\frac{30}{r^{2}}D_{\left[ 3,4,0\right] }%
\end{array}%
\right) $

------------------------------------------------------------------------------------------------------

\begin{equation*}
N=8
\end{equation*}

$\left( \bigskip 
\begin{array}{c}
\left( \partial +\frac{2}{r}\right) D_{\left[ 9,0,0\right] }-\frac{16}{r}D_{%
\left[ 7,2,0\right] } \\ 
\left( \partial +\frac{4}{r}\right) D_{\left[ 7,2,0\right] }-\frac{8}{r}D_{%
\left[ 5,4,0\right] } \\ 
\left( \partial +\frac{6}{r}\right) D_{\left[ 5,4,0\right] }-\frac{24}{5r}D_{%
\left[ 3,6,0\right] } \\ 
\left( \partial +\frac{8}{r}\right) D_{\left[ 3,6,0\right] }-\frac{16}{7r}D_{%
\left[ 1,8,0\right] } \\ 
\left( \partial +\frac{10}{r}\right) D_{\left[ 1,8,0\right] }%
\end{array}%
\right) $ $\ \left( 
\begin{array}{c}
\left( \partial _{r}^{2}+\frac{2}{r}\partial _{r}-\frac{16}{r^{2}}\right)
D_{8,0,0}+\frac{112}{r^{2}}D_{6,2,0} \\ 
\left( \partial _{r}^{2}+\frac{2}{r}\partial _{r}-\frac{38}{r^{2}}\right)
D_{6,2,0}+\frac{2}{r^{2}}D_{8,0,0}+\frac{40}{r^{2}}D_{4,4,0} \\ 
\left( \partial _{r}^{2}+\frac{2}{r}\partial _{r}-\frac{44}{r^{2}}\right)
D_{4,4,0}+\frac{12}{r^{2}}D_{6,2,0}+\frac{72}{5r^{2}}D_{2,6,0} \\ 
\left( \partial _{r}^{2}+\frac{2}{r}\partial _{r}-\frac{34}{r^{2}}\right)
D_{2,6,0}+\frac{30}{r^{2}}D_{4,4,0}+\frac{16}{7r^{2}}D_{0,8,0} \\ 
\left( \partial _{r}^{2}+\frac{2}{r}\partial _{r}-\frac{8}{r^{2}}\right)
D_{0,8,0}+\frac{56}{r^{2}}D_{2,6,0}%
\end{array}%
\right) $

------------------------------------------------------------------------------------------------------.

\begin{acknowledgement}
The author thanks Mr. Jonas Boschung and Prof. Norbert Peters for
significant help correcting perviously incorrect Laplacian operators. \ The
author thanks the organizers of the Hydrodynamics Turbulence Program held at
the Institute for Theoretical Physics, University of California at Santa
Barbara, whereby this research was supported in part by the National Science
Foundation under grant number PHY94-07194.
\end{acknowledgement}

\section{REFERENCES}

Abramowitz, M. and I. A. Stegun 1964 \textit{Handbook of Mathematical
Functions with Formulas, Graphs, and Mathematical Tables, National Bureau of
Standards Applied Mathematics Series 55}, U. S. Government Printing Office,
Washington DC.

Hill, R. J. 1997 Applicability of Kolmogorov's and Monin's equations of
turbulence. \textit{J. Fluid Mech.} \textbf{353}, 67.

Hill, R. J. and O. N. Boratav 2001 Next-order structure-function equations. 
\textit{Phys. Fluids} \textbf{13}, 276.

Millionshtchikov, M. D. 1941 On the theory of homogeneous isotropic
turbulence. \textit{Dokl. Akad. Nauk. SSSR} \textbf{32}, 611.

\section{APPENDIX A: \ The viscous term}

\qquad The quantity $\left\{ v_{j}v_{k}\cdot \cdot \cdot v_{l}\partial
_{x_{n}}\partial _{x_{n}}v_{i}\right\} $\ requires special attention. \
Consider the repeated application of the identity 
\begin{equation}
\partial _{x_{n}}\partial _{x_{n}}\left( fg\right) =f\partial
_{x_{n}}\partial _{x_{n}}g+g\partial _{x_{n}}\partial _{x_{n}}f+2\left(
\partial _{x_{n}}f\right) \left( \partial _{x_{n}}g\right)
\label{1Laplaceident}
\end{equation}
to the quantity 
\begin{equation}
\partial _{x_{n}}\partial _{x_{n}}\left( v_{j}v_{k}v_{m}\cdot \cdot \cdot
v_{i}\right)  \label{1testexpression}
\end{equation}
for $N$ factors of velocity difference in $\left( v_{j}v_{k}v_{m}\cdot \cdot
\cdot v_{i}\right) $. \ For the first application of (\ref{1Laplaceident})
let $f=v_{j}$ and let $g$ be the remaining factors $\left( v_{k}v_{m}\cdot
\cdot \cdot v_{i}\right) $; this gives 
\begin{equation}
\partial _{x_{n}}\partial _{x_{n}}\left( v_{j}v_{k}v_{m}\cdot \cdot \cdot
v_{i}\right) =v_{j}\partial _{x_{n}}\partial _{x_{n}}\left( v_{k}v_{m}\cdot
\cdot v_{i}\right) +\left( v_{k}v_{m}\cdot \cdot v_{i}\right) \partial
_{x_{n}}\partial _{x_{n}}v_{j}+2\left[ \partial _{x_{n}}v_{j}\right] \left[
\partial _{x_{n}}\left( v_{k}v_{m}\cdot \cdot v_{i}\right) \right] .
\label{1Lapreduc}
\end{equation}
\ From the differentiation chain rule, $\partial _{x_{n}}\left(
v_{k}v_{m}\cdot \cdot v_{i}\right) $ is the sum of $N-1$ terms of the form $%
\left( v_{m}\cdot \cdot v_{p}\partial _{x_{n}}v_{i}\right) $. \ Thus, the
right-most term in (\ref{1Lapreduc}) is $N-1$ terms of the form $2v_{m}\cdot
\cdot \cdot v_{p}\left( \partial _{x_{n}}v_{i}\right) \left( \partial
_{x_{n}}v_{j}\right) $ each term containing $N$ factors; two of those
factors are distinguished by being derivatives of velocity differences. \
The second application of (\ref{1Laplaceident}) is performed on $%
v_{j}\partial _{x_{n}}\partial _{x_{n}}\left( v_{k}v_{m}\cdot \cdot
v_{i}\right) $\ in (\ref{1Lapreduc}), for which purpose $f=v_{k}$ and $%
g=\left( v_{m}\cdot \cdot v_{i}\right) $; this gives 
\begin{equation*}
v_{j}\partial _{x_{n}}\partial _{x_{n}}\left( v_{k}v_{m}\cdot \cdot \cdot
v_{i}\right) =v_{j}v_{k}\partial _{x_{n}}\partial _{x_{n}}\left( v_{m}\cdot
\cdot v_{i}\right) +v_{j}\left( v_{m}\cdot \cdot v_{i}\right) \partial
_{x_{n}}\partial _{x_{n}}v_{k}+2v_{j}\left[ \partial _{x_{n}}v_{k}\right] %
\left[ \partial _{x_{n}}\left( v_{m}\cdot \cdot v_{i}\right) \right] .
\end{equation*}
The right-most term gives $N-2$ terms of the form $2v_{j}v_{m}\cdot \cdot
\cdot v_{p}\left( \partial _{x_{n}}v_{i}\right) \left( \partial
_{x_{n}}v_{k}\right) $ each term containing $N$ factors.

\qquad There are $N-1$ steps to complete reduction of the formula. \ The
number of terms of the form $2v_{j}v_{m}\cdot \cdot \cdot v_{p}\left(
\partial _{x_{n}}v_{i}\right) \left( \partial _{x_{n}}v_{k}\right) $ is $%
\left( N-1\right) $ from the first step, $\left( N-2\right) $ from the
second step, etc. such that the total number of terms is $\left( N-1\right)
+\left( N-2\right) +\cdot \cdot \cdot +\left( N-\left( N-1\right) \right)
=N\left( N-1\right) /2$. \ Now, $N\left( N-1\right) /2=\binom{N}{2}$ is the
binomial coefficient equal to the number of ways of choosing two indexes
from a set of $N$\ indexes; the quantities $\left( \partial
_{x_{n}}v_{i}\right) $ and $\left( \partial _{x_{n}}v_{j}\right) $\ in $%
2v_{m}\cdot \cdot \cdot v_{p}\left( \partial _{x_{n}}v_{i}\right) \left(
\partial _{x_{n}}v_{j}\right) $\ contain the chosen two indexes $i$\ and $j$%
. \ The $\binom{N}{2}$\ terms constitute $2\left\{ v_{m}\cdot \cdot \cdot
v_{p}\left( \partial _{x_{n}}v_{i}\right) \left( \partial
_{x_{n}}v_{j}\right) \right\} $. Because two factors of the form $\left(
v_{j}v_{m}\cdot \cdot v_{i}\right) \partial _{x_{n}}\partial _{x_{n}}v_{k}$
appear in the last step, the total number of terms of the form $\left(
v_{j}\cdot \cdot \cdot v_{n}\right) \partial _{x_{n}}\partial _{x_{n}}v_{i}$%
\ is $N$. \ Not surprisingly, these $N$\ terms constitute $\left\{ \left(
v_{j}\cdot \cdot \cdot v_{n}\right) \partial _{x_{n}}\partial
_{x_{n}}v_{i}\right\} $, and $N=\binom{N}{1}$ is the binomial coefficient
equal to the number of ways of choosing one index from a set of $N$\
indexes, the quantity $\partial _{x_{n}}\partial _{x_{n}}v_{i}$\ contains
the chosen one index $i$.

That is, for any $N$%
\begin{equation}
\partial _{x_{n}}\partial _{x_{n}}\left( v_{j}\cdot \cdot \cdot v_{i}\right)
=\left\{ \left( v_{j}\cdot \cdot \cdot v_{l}\right) \partial
_{x_{n}}\partial _{x_{n}}v_{i}\right\} +2\left\{ v_{k}\cdot \cdot \cdot
v_{l}\left( \partial _{x_{n}}v_{i}\right) \left( \partial
_{x_{n}}v_{j}\right) \right\} .  \label{1regslogident}
\end{equation}
The left-hand side is symmetric under interchange of any pair of indexes
(not including $n$ because summation is implied over $n$), and the French
brackets make the right-hand side likewise symmetric.

\qquad Use of (\ref{1regslogident}) within the viscous term $\nu \left\{
v_{j}v_{k}\cdot \cdot \cdot v_{l}\left( \partial _{x_{n}}\partial
_{x_{n}}v_{i}+\partial _{x_{n}^{\prime }}\partial _{x_{n}^{\prime
}}v_{i}\right) \right\} $ that arrises from (\ref{2ndstep}), gives 
\begin{eqnarray}
&&\left\{ v_{j}v_{k}\cdot \cdot \cdot v_{l}\left( \partial _{x_{n}}\partial
_{x_{n}}v_{i}+\partial _{x_{n}^{\prime }}\partial _{x_{n}^{\prime
}}v_{i}\right) \right\}  \notag \\
&=&\left( \partial _{x_{n}}\partial _{x_{n}}+\partial _{x_{n}^{\prime
}}\partial _{x_{n}^{\prime }}\right) \left( v_{j}\cdot \cdot \cdot
v_{i}\right)  \notag \\
&&-2\left\{ v_{k}\cdot \cdot \cdot v_{l}\left[ \left( \partial
_{x_{n}}v_{i}\right) \left( \partial _{x_{n}}v_{j}\right) +\left( \partial
_{x_{n}^{\prime }}v_{i}\right) \left( \partial _{x_{n}^{\prime
}}v_{j}\right) \right] \right\} ,  \label{1VISCOUS}
\end{eqnarray}%
where the right-most term in (\ref{1regslogident}) has been subtracted from
both sides of (\ref{1regslogident}) to obtain (\ref{1VISCOUS}). \ Note that $%
\left( \partial _{x_{n}}u_{i}\right) \left( \partial _{x_{n}}u_{j}\right)
=\left( \partial _{x_{n}}v_{i}\right) \left( \partial _{x_{n}}v_{j}\right) $
and $\left( \partial _{x_{n}^{\prime }}u_{i}^{\prime }\right) \left(
\partial _{x_{n}^{\prime }}u_{j}^{\prime }\right) =\left( \partial
_{x_{n}^{\prime }}v_{i}\right) \left( \partial _{x_{n}^{\prime
}}v_{j}\right) $, and that use of (\ref{1identderivs}) gives 
\begin{equation*}
\left( \partial _{x_{n}}\partial _{x_{n}}+\partial _{x_{n}^{\prime
}}\partial _{x_{n}^{\prime }}\right) =2\left( \partial _{r_{n}}\partial
_{r_{n}}+\frac{1}{4}\partial _{X_{n}}\partial _{X_{n}}\right) .
\end{equation*}%
Then, (\ref{1VISCOUS}) can be written as 
\begin{eqnarray}
&&\left\{ v_{j}v_{k}\cdot \cdot \cdot v_{l}\left( \partial _{x_{n}}\partial
_{x_{n}}v_{i}+\partial _{x_{n}^{\prime }}\partial _{x_{n}^{\prime
}}v_{i}\right) \right\}  \notag \\
&=&2\left( \partial _{r_{n}}\partial _{r_{n}}+\frac{1}{4}\partial
_{X_{n}}\partial _{X_{n}}\right) \left( v_{j}\cdot \cdot \cdot v_{i}\right) 
\notag \\
&&-2\left\{ v_{k}\cdot \cdot \cdot v_{l}\left[ \left( \partial
_{x_{n}}u_{i}\right) \left( \partial _{x_{n}}u_{j}\right) +\left( \partial
_{x_{n}^{\prime }}u_{i}^{\prime }\right) \left( \partial _{x_{n}^{\prime
}}u_{j}^{\prime }\right) \right] \right\} .
\end{eqnarray}

\section{APPENDIX B: \ Derivation of (14)}

\qquad The purpose of this appendix is to derive (\ref{1Palternative}). \
Since (\ref{1identderivs}) allows $P_{i}$\ to be written as $P_{i}=\partial
_{X_{i}}\left( p-p^{\prime }\right) $, the differentiation chain rule gives 
\begin{equation}
v_{j}v_{k}\cdot \cdot \cdot v_{l}P_{i}=v_{j}v_{k}\cdot \cdot \cdot
v_{l}\partial _{X_{i}}\left( p-p^{\prime }\right) =\partial _{X_{i}}\left[
v_{j}v_{k}\cdot \cdot \cdot v_{l}\left( p-p^{\prime }\right) \right] -\left(
p-p^{\prime }\right) \left\{ \left( \partial _{X_{i}}v_{j}\right) v_{k}\cdot
\cdot \cdot v_{l}\right\} _{/i/},  \label{1Peqn}
\end{equation}%
where the notation $\left\{ \circ \right\} _{/i/}$ denotes the sum of all
terms of a given type that produce symmetry under interchange of each pair
of indexes with the index $i$ excluded. \ Recall that the product $%
v_{j}v_{k}\cdot \cdot \cdot v_{l}$ consists of $N-1$ factors. \ Sum the $N$\
equations of type (\ref{1Peqn}) such that the sum is even under interchange
of all pairs of indexes; then 
\begin{equation}
\left\{ v_{j}v_{k}\cdot \cdot \cdot v_{l}P_{i}\right\} =\left\{ \partial
_{X_{i}}\left[ v_{j}v_{k}\cdot \cdot \cdot v_{l}\left( p-p^{\prime }\right) %
\right] \right\} -\left( N-1\right) \left( p-p^{\prime }\right) \left\{
\left( \partial _{X_{i}}v_{j}\right) v_{k}\cdot \cdot \cdot v_{l}\right\} ,
\label{1Presult}
\end{equation}%
where use was made of the fact that the $N-1$ terms in the sum $\left\{
\left( \partial _{X_{i}}v_{j}\right) v_{k}\cdot \cdot \cdot v_{l}\right\}
_{/i/}$\ each give the same result, namely, $\left\{ \left( \partial
_{X_{i}}v_{j}\right) v_{k}\cdot \cdot \cdot v_{l}\right\} $. \ From (\ref%
{1identderivs}), $\partial _{X_{i}}v_{j}=\partial _{x_{i}}u_{j}-\partial
_{x_{j}^{\prime }}u_{i}^{\prime }$; such that the definition of strain rate (%
\ref{1straindefine}) gives 
\begin{equation}
\left( \partial _{X_{i}}v_{j}+\partial _{X_{j}}v_{i}\right)
/2=s_{ij}-s_{ij}^{\prime }.  \label{2indentderivs}
\end{equation}%
Use of (\ref{2indentderivs})\ gives $\left\{ \left( \partial
_{X_{i}}v_{j}\right) v_{k}\cdot \cdot \cdot v_{l}\right\} =2\left\{ \left(
s_{ij}-s_{ij}^{\prime }\right) v_{k}\cdot \cdot \cdot v_{l}\right\} $,
substitution of which into (\ref{1Presult}) gives 
\begin{equation}
\left\{ v_{j}v_{k}\cdot \cdot \cdot v_{l}P_{i}\right\} =\left\{ \partial
_{X_{i}}\left[ v_{j}v_{k}\cdot \cdot \cdot v_{l}\left( p-p^{\prime }\right) %
\right] \right\} -2\left( N-1\right) \left( p-p^{\prime }\right) \left\{
\left( s_{ij}-s_{ij}^{\prime }\right) v_{k}\cdot \cdot \cdot v_{l}\right\} .
\label{1AlterZ}
\end{equation}

Note that the factor of $2$\ can be verified by the following sample
calculation%
\begin{eqnarray*}
\left\{ \left( \partial _{X_{i}}v_{j}\right) v_{k}\right\}  &=&\left(
\partial _{X_{i}}v_{j}\right) v_{k}+\left( \partial _{X_{j}}v_{i}\right)
v_{k}+\left( \partial _{X_{k}}v_{j}\right) v_{i}+\left( \partial
_{X_{j}}v_{k}\right) v_{i}+\left( \partial _{X_{i}}v_{k}\right) v_{j}+\left(
\partial _{X_{k}}v_{i}\right) v_{j} \\
&=&\left[ \partial _{X_{i}}v_{j}+\partial _{X_{j}}v_{i}\right] v_{k}+\left[
\partial _{X_{k}}v_{j}+\partial _{X_{j}}v_{k}\right] v_{i}+\left[ \partial
_{X_{i}}v_{k}+\partial _{X_{k}}v_{i}\right] v_{j} \\
&=&2\left( \left[ s_{ij}-s_{ij}^{\prime }\right] v_{k}+\left[
s_{kj}-s_{kj}^{\prime }\right] v_{i}+\left[ s_{ik}-s_{ik}^{\prime }\right]
v_{j}\right)  \\
&=&2\left\{ \left( s_{ij}-s_{ij}^{\prime }\right) v_{k}\right\} 
\end{eqnarray*}

\section{APPENDIX C: \ The Coefficient in (21)}

\qquad The purpose of this appendix is to obtain a formula for evaluation of
the coefficient $\left\{ \mathbf{W}_{\left[ N-2P\right] }\mathbf{\delta }_{%
\left[ 2P\right] }\right\} $\ in the special Cartesian coordinate system. \
In this coordinate system, $\frac{r_{i}}{r}=\delta _{1i}$ such that $\mathbf{%
W}_{\left[ N\right] }$ is the product of $N$ Kronecker deltas of the form $%
\delta _{1i}$. \ Consider setting $N_{1}$\ of the indexes equal to $1$, an
even number $N_{2}$\ to $2$, and an even number $N_{3}$\ to $3$ such that $%
N=N_{1}+N_{2}+N_{3}$. \ Then $\left\{ \mathbf{W}_{\left[ N-2P\right] }%
\mathbf{\delta }_{\left[ 2P\right] }\right\} $ becomes a sum of zeros and
ones. \ What is that sum? \ If all indexes are set to $1$, i.e., $N_{1}=N$,
then all terms in the sum $\left\{ \mathbf{W}_{\left[ N-2P\right] }\mathbf{%
\delta }_{\left[ 2P\right] }\right\} $ are unity such that $\left\{ \mathbf{W%
}_{\left[ N-2P\right] }\mathbf{\delta }_{\left[ 2P\right] }\right\} $\
equals its number of terms; from (\ref{numbterms}) in Appendix E, that
number is $\binom{N}{N-2P}\left( 2P-1\right) !!$. \ The notation $\binom{N}{%
N-2P}$ is a binomial coefficient. \ To interpret $\left( 2P-1\right) !!$,
recall that $q!!\equiv q\left( q-2\right) \left( q-4\right) \cdot \cdot
\cdot q_{L}$, \ where $q_{L}$ is $2$ or $1$ for $q$ even or odd,
respectively, and $\left( -1\right) !!\equiv 1$. \ Now consider setting two
indexes to $2$, thus $N_{1}=N-2$, and $N_{2}=2$. \ Name the two indexes $i=2$
and $j=2$. \ The only term in $\left\{ \mathbf{W}_{\left[ N-2P\right] }%
\mathbf{\delta }_{\left[ 2P\right] }\right\} $\ that is nonzero is that
which has $i$ and $j$ together in a single Kronecker delta, $\delta _{ij}$,
within $\mathbf{\delta }_{\left[ 2P\right] }$. \ For $P=0$ there is no such $%
\delta _{ij}$, in which case\ $\left\{ \mathbf{W}_{\left[ N\right] }\mathbf{%
\delta }_{\left[ 0\right] }\right\} =0$. \ For $P\geq 1$, there is one such $%
\delta _{ij}$, which is set to $1$ and it multiplies the quantity $\left\{ 
\mathbf{W}_{\left[ N-2P\right] }\mathbf{\delta }_{\left[ 2\left( P-1\right) %
\right] }\right\} $; since this quantity is evaluated with all $1$s, it is
equal to its number of terms, namely $\binom{\left( N-2P\right) +2\left(
P-1\right) }{\left( N-2P\right) }\left( 2\left( P-1\right) -1\right) !!$. \
Now consider setting four indexes to $2$; thus $N_{1}=N-4$, and $N_{2}=4$. \
Name the four indexes $i=2$, $j=2$, $k=2$, $l=2$. \ The only terms in $%
\left\{ \mathbf{W}_{\left[ N-2P\right] }\mathbf{\delta }_{\left[ 2P\right]
}\right\} $\ that are nonzero are those that have factors $\delta
_{ij}\delta _{kl}$, $\delta _{ik}\delta _{jl}$, or $\delta _{il}\delta _{jk}$
within $\mathbf{\delta }_{\left[ 2P\right] }$. \ For $P\leq 1$ there is no
such pair of Kronecker deltas such that\ $\left\{ \mathbf{W}_{\left[ N\right]
}\mathbf{\delta }_{\left[ 0\right] }\right\} =0$ and\ $\left\{ \mathbf{W}_{%
\left[ N-2\right] }\mathbf{\delta }_{\left[ 2\right] }\right\} =0$. \ For $%
P\geq 2$, there are the above $3=\left( N_{2}-1\right) !!$ nonzero factors
and each multiplies the quantity $\left\{ \mathbf{W}_{\left[ N-2P\right] }%
\mathbf{\delta }_{\left[ 2\left( P-2\right) \right] }\right\} $; since this
latter quantity is subsequently evaluated with all $1$s, it is equal to its
number of terms, namely $\binom{\left( N-2P\right) +2\left( P-2\right) }{%
\left( N-2P\right) }\left( 2\left( P-2\right) -1\right) !!$. \ Continuation
of this study reveals the pattern that when $\left\{ \mathbf{W}_{\left[ N-2P%
\right] }\mathbf{\delta }_{\left[ 2P\right] }\right\} $ is evaluated with $%
N_{2}$ $2$s and $N_{1}$ $1$s such that $N=N_{1}+N_{2}$ then 
\begin{eqnarray*}
\text{if }2P &<&N_{2}\text{ then }\left\{ \mathbf{W}_{\left[ N-2P\right] }%
\mathbf{\delta }_{\left[ 2P\right] }\right\} =0,\text{ otherwise} \\
\left\{ \mathbf{W}_{\left[ N-2P\right] }\mathbf{\delta }_{\left[ 2P\right]
}\right\} &=&\left( N_{2}-1\right) !!\binom{\left( N-2P\right) +2\left( P-%
\frac{N_{2}}{2}\right) }{\left( N-2P\right) }\left( 2\left( P-\frac{N_{2}}{2}%
\right) -1\right) !!. \\
&=&\left( N_{2}-1\right) !!\left[ \frac{N_{1}!}{\left( N_{1}-2\left( P-\frac{%
N_{2}}{2}\right) \right) !\left( 2\left( P-\frac{N_{2}}{2}\right) \right) !}%
\right] \left( 2\left( P-\frac{N_{2}}{2}\right) -1\right) !!
\end{eqnarray*}%
If one ceases increasing the number of $2$s and commences increasing the
number of $3$s in pairs such that $N=N_{1}+N_{2}+N_{3}$, then $\left(
N_{2}-1\right) !!\binom{\left( N-2P\right) +2\left( P-\frac{N_{2}}{2}\right) 
}{\left( N-2P\right) }\left( 2\left( P-\frac{N_{2}}{2}\right) -1\right) !!$
is replaced by $\left( N_{2}-1\right) !!\left( N_{3}-1\right) !!\binom{%
\left( N-2P\right) +2\left( P-\frac{N_{2}}{2}-\frac{N_{3}}{2}\right) }{%
\left( N-2P\right) }\left( 2\left( P-\frac{N_{2}}{2}-\frac{N_{3}}{2}\right)
-1\right) !!$. \ Of course, the binomial coefficient can be expressed as
follows: $\binom{\left( N-2P\right) +2\left( P-\frac{N_{2}}{2}-\frac{N_{3}}{2%
}\right) }{\left( N-2P\right) }=\left( N-N_{2}-N_{3}\right) !/\left[ \left(
N-2P\right) !\left( 2P-N_{2}-N_{3}\right) !\right] $; also, $\left(
N-N_{2}-N_{3}\right) !=N_{1}!$. \ The double factorial can be eliminated by
means of the following identities: 
\begin{equation}
\left( 2Q-1\right) !!/\left( 2Q\right) !=1/\left( 2Q\right) !!=1/\left(
2^{Q}Q!\right) .  \label{doublefacident}
\end{equation}%
That is, 
\begin{eqnarray*}
&&\left( 2P-N_{2}-N_{3}-1\right) !!/\left( 2P-N_{2}-N_{3}\right) ! \\
&=&1/\left( 2P-N_{2}-N_{3}\right) !!=1/\left[ 2^{\left( P-\frac{N_{2}}{2}-%
\frac{N_{3}}{2}\right) }\left( P-\frac{N_{2}}{2}-\frac{N_{3}}{2}\right) !%
\right] .
\end{eqnarray*}%
Also, $\left( N_{2}-1\right) !!=N_{2}!/\left[ 2^{N_{2}/2}\left(
N_{2}/2\right) !\right] $. \ Finally, 
\begin{eqnarray}
\text{if }2P &<&N_{2}+N_{3}\text{ then }\left\{ \mathbf{W}_{\left[ N-2P%
\right] }\mathbf{\delta }_{\left[ 2P\right] }\right\} =0,\text{ otherwise}
\label{AppC1} \\
\left\{ \mathbf{W}_{\left[ N-2P\right] }\mathbf{\delta }_{\left[ 2P\right]
}\right\} &=&\left( N_{2}-1\right) !!\left( N_{3}-1\right) !!N_{1}!/\left[
\left( N-2P\right) !2^{\left( P-\frac{N_{2}}{2}-\frac{N_{3}}{2}\right)
}\left( P-\frac{N_{2}}{2}-\frac{N_{3}}{2}\right) !\right]  \notag \\
&=&N_{1}!N_{2}!N_{3}!/\left[ \left( N-2P\right) !2^{P}\left( \frac{N_{2}}{2}%
\right) !\left( \frac{N_{3}}{2}\right) !\left( P-\frac{N_{2}}{2}-\frac{N_{3}%
}{2}\right) !\right] .  \label{AppC2}
\end{eqnarray}

\section{APPENDIX D: \ Properties of Isotropic Symmetric Tensors}

\qquad The purpose of this appendix is to determine (1) which components of
a symmetric, isotropic tensor are zero; (2) how many components are zero and
how many are nonzero; and (3) the relationships between the nonzero
components.

\qquad Consider which\ components $D_{\left[ N:N_{1},N_{2},N_{3}\right] }$
are nonzero and which are zero. \ In (\ref{1isoformula}), $\mathbf{W}_{\left[
N-2P\right] }$ vanishes if any of its indexes is $2$ or $3$, $\mathbf{\delta 
}_{\left[ 2P\right] }$ vanishes unless it contains an even number of indexes
equal to $2$ and a likewise even number of $3$s (and of $1$s). \ Thus, a
component $D_{\left[ N:N_{1},N_{2},N_{3}\right] }$ is nonzero only if both $%
N_{2}$ and $N_{3}$ are even. \ Because $N=N_{1}+N_{2}+N_{3}$, $D_{\left[
N:N_{1},N_{2},N_{3}\right] }$ is nonzero only when $N_{1}$ is odd if $N$ is
odd and only when $N_{1}$ is even if $N$ is even. \ The values of $N_{1}$
that can give nonzero values of $D_{\left[ N:N_{1},N_{2},N_{3}\right] }$ are 
$N$, $N-2$, $\cdot \cdot \cdot $, $0$ or $1$; i.e., $M+1$ cases of$\ N_{1}$.
\ Given $N_{1}$, the values of $\left[ N_{2},N_{3}\right] $ that give
nonzero values of $D_{\left[ N:N_{1},N_{2},N_{3}\right] }$ are $\left[
N-N_{1},0\right] $, $\left[ N-N_{1}-2,2\right] $, $\cdot \cdot \cdot $,$%
\left[ 0,N-N_{1}\right] $; i.e., $\left( N-N_{1}+2\right) /2$ cases (note
that $N-N_{1}=N_{2}+N_{3}$ is necessarily even). \ Counting the number of
cases of $\left[ N_{2},N_{3}\right] $ as $N_{1}$ varies from $N$\ to $0$\
or\ $1$, (i.e., substituting $\ N_{1}=N$, then $N_{1}=N-2$, $\cdot \cdot
\cdot $ into $\left( N-N_{1}+2\right) /2$ and adding the resultant numbers)
shows that there are $1+2+3+\cdot \cdot \cdot +\left( M+1\right) =\left(
M+1\right) \left( M+2\right) /2$\ components $D_{\left[ N:N_{1},N_{2},N_{3}%
\right] }$\ that are nonzero. \ Since there are $3^{N}$ components of $%
\mathbf{D}_{\left[ N\right] }$, the remaining $3^{N}-\left( M+1\right)
\left( M+2\right) /2$ components are zero. \ Since there are $M+1$\ linearly
independent components (that are related to the $D_{N,P}$), there are $%
\left( M+1\right) \left( M+2\right) /2-\left( M+1\right) =M\left( M+1\right)
/2$\ relationships among the nonzero components $D_{\left[
N:N_{1},N_{2},N_{3}\right] }.$\ \ For instance,\ interchange of the values
of $N_{2}$ and $N_{3}$ produces components that are equal.

\qquad Consider the $M+1$\ linearly independent equations that determine\
the $D_{N,P}$ in terms of $M+1$ of the $D_{\left[ N:N_{1},N_{2},N_{3}\right]
}$. \ $D_{N,0}$ is related by (\ref{1isoformula}) to only the component $D_{%
\left[ N:N,0,0\right] }$ because (\ref{0sinisocoefs}) shows that the
coefficient of $D_{N,0}$, namely $\left\{ \mathbf{W}_{\left[ N\right] }%
\mathbf{\delta }_{\left[ 0\right] }\right\} $,\ vanishes unless $N_{1}=N$. \
That is, if all indexes in (\ref{1isoformula}) are $1$, then $D_{\left[
N:N,0,0\right] }$ appears on the left-hand side and the $D_{N,P}$\ for all $%
P $ appear in the equation. \ This equation is essential for determining $%
D_{N,0} $ and is called ``the equation for $D_{N,0}$''; similar terminology
``the equation for $D_{N,P}$'' is used below. \ With the equation for $%
D_{N,0}$ in hand, consider $D_{N,1}$. \ $D_{N,1}$ is related by (\ref%
{1isoformula}) to $D_{\left[ N:N-2,2,0\right] }$\ or $D_{\left[ N:N-2,0,2%
\right] }$. \ When $D_{\left[ N:N-2,2,0\right] }$\ or $D_{\left[ N:N-2,0,2%
\right] }$\ is on the left-hand side of (\ref{1isoformula}) the equation for 
$D_{N,1}$\ results because the coefficients $\left\{ \mathbf{W}_{\left[ N-2P%
\right] }\mathbf{\delta }_{\left[ 2P\right] }\right\} $\ of $D_{N,P}$ for $%
P\geq 1$ do not vanish, but the coefficient of $D_{N,0}$ does vanish. \ Now
consider an equation for $D_{N,2}$. \ $D_{N,2}$ is related by (\ref%
{1isoformula}) to $D_{\left[ N:N-4,4,0\right] }$\ or $D_{\left[ N:N-4,0,4%
\right] }$ or $D_{\left[ N:N-4,2,2\right] }$; these components also involve $%
D_{N,P}$ for $P\geq 3$ but not for $P\leq 1$. \ This procedure repeats until
the last equation is produced; only $D_{N,M}$\ appears in the last equation.
\ If $N$ is even then $D_{N,M}$ is related to $D_{\left[ N:0,N_{2},N_{3}%
\right] }$ with $N_{2} $ and $N_{3}$ equal to any positive even numbers such
that $N=N_{2}+N_{3}$. \ If $N$ is odd then $D_{N,M}$ is related to $D_{\left[
N:1,N_{2},N_{3}\right] }$ with $N_{2}$ and $N_{3}$ equal to any positive
even numbers such that $N=1+N_{2}+N_{3}$. \ This procedure results in a set
of $M+1$ linearly independent equations that can be solved to obtain the $%
D_{N,P}$\ in terms of the $D_{\left[ N:N_{1},N_{2},N_{3}\right] }$. \ Note
that $M+1$ components must be chosen for use in the $M+1$ equations. \ For
instance, from the above example, $D_{\left[ N:N,0,0\right] }$\ must be
used; either $D_{\left[ N:N-2,2,0\right] }$\ or $D_{\left[ N:N-2,0,2\right]
} $ must be chosen, and one of $D_{\left[ N:N-4,4,0\right] }$\ or $D_{\left[
N:N-4,0,4\right] }$ or $D_{\left[ N:N-4,2,2\right] }$\ must be chosen, etc.
\ For simplicity, the chosen components can all have $N_{3}=0$, i.e., the
choice can be $D_{\left[ N:N,0,0\right] }$, $D_{\left[ N:N-2,2,0\right] }$, $%
D_{\left[ N:N-4,4,0\right] }$, $\cdot \cdot \cdot $, $D_{\left[ N:N-2M,2M,0%
\right] }$.

\qquad The above procedure also reveals algebraic relationships between the
nonzero $D_{\left[ N:N_{1},N_{2},N_{3}\right] }$. \ The equation for $%
D_{N,1} $\ can be expressed in terms of either $D_{\left[ N:N-2,2,0\right] }$%
\ or $D_{\left[ N:N-2,0,2\right] }$; the left-hand side is the same in
either case because the coefficients (\ref{1isocoefs}) are the same; hence $%
D_{\left[ N:N-2,2,0\right] }=D_{\left[ N:N-2,0,2\right] }$. \ The equation
for $D_{N,2} $\ can be expressed in terms of $D_{\left[ N:N-4,2,2\right] }$
or $D_{\left[ N:N-4,4,0\right] }$\ or $D_{\left[ N:N-4,0,4\right] }$ such
that (\ref{1isocoefs}) gives $D_{\left[ N:N-4,4,0\right] }=D_{\left[
N:N-4,0,4\right] } $; but what is the relationship of $D_{\left[ N:N-4,2,2%
\right] }$ to $D_{\left[ N:N-4,4,0\right] }$ and $D_{\left[ N:N-4,0,4\right]
}$? \ When $D_{\left[ N:N-4,0,4\right] }$ or $D_{\left[ N:N-4,4,0\right] }$
is on the left-hand side of (\ref{1isoformula}), the nonzero coefficients
are, from (\ref{1isocoefs}), $\left( N-4\right) !4!0!/\left[ \left(
N-2P\right) !2^{P}2!0!\left( P-2-0\right) !\right] $, but when $D_{\left[
N:N-4,2,2\right] }$ is on the left-hand side, the nonzero coefficients are $%
\left( N-4\right) !2!2!/\left[ \left( N-2P\right) !2^{P}1!1!\left(
P-1-1\right) !\right] $. \ The ratio of these coefficients is $3$. \ Since
this ratio is independent of $P$, the entire right-hand side of (\ref%
{1isoformula}) is three times greater when $D_{\left[ N:N-4,0,4\right] }$\
is on the left-hand side as compared to when $D_{\left[ N:N-4,2,2\right] }$\
is on the left-hand side. \ Therefore, the proportionality $D_{\left[
N:N-4,0,4\right] }:D_{\left[ N:N-4,2,2\right] }$ (and also $D_{\left[
N:N-4,4,0\right] }:D_{\left[ N:N-4,2,2\right] }$) is $3:1$. \ In general,
for given $N$ and $N_{1}$, and hence given $N_{2}+N_{3}=N-N_{1}$, and
another choice of $N_{2}$ and $N_{3}$, call them $N_{2}^{\prime }$ and $%
N_{3}^{\prime }$, such that $N_{2}^{\prime }+N_{3}^{\prime
}=N_{2}+N_{3}=N-N_{1}$, the proportionality obtained from (\ref{1isocoefs})
is $D_{\left[ N:N_{1},N_{2},N_{3}\right] }:D_{\left[ N:N_{1},N_{2}^{\prime
},N_{3}^{\prime }\right] }=\left[ N_{2}!N_{3}!/\left( N_{2}/2\right) !\left(
N_{3}/2\right) !\right] :\left[ N_{2}^{\prime }!N_{3}^{\prime }!/\left(
N_{2}^{\prime }/2\right) !\left( N_{3}^{\prime }/2\right) !\right] $.$\ \ $%
Parameterized in terms of an integer $L$ such that $N=N_{1}+2L$, for given $%
N_{1}$, the proportionalities are $D_{\left[ N:N_{1},2L,0\right] }:D_{\left[
N:N_{1},2L-2,2\right] }:D_{\left[ N:N_{1},2L-4,4\right] }:\cdot \cdot \cdot
:D_{\left[ N:N_{1},0,2L\right] }=\left[ \left( 2L\right) !0!/L!0!\right] :%
\left[ \left( 2L-2\right) !2!/\left( L-1\right) !1!\right] :\left[ \left(
2L-4\right) !4!/\left( L-2\right) !2!\right] :\cdot \cdot \cdot :\left[
0!\left( 2L\right) !/0!L!\right] $. \ This constitutes $\left(
N-N_{1}\right) /2$ relationships among the nonzero components $D_{\left[
N:N_{1},N_{2},N_{3}\right] }$\ for given $N_{1}$. \ Substituting the $M+1$
cases of $N_{1}$ (i.e., $N_{1}=N$, $N_{1}=N-1$, $\cdot \cdot \cdot $) into $%
\left( N-N_{1}\right) /2$\ the number of relationships thus identified among
the components of $\mathbf{D}_{\left[ N\right] }$ is $0+1+2+\cdot \cdot
\cdot +M=M\left( M+1\right) /2$. \ In the paragraph above, it was determined
that the total number of relationships among the nonzero components of $%
\mathbf{D}_{\left[ N\right] }$ is $M\left( M+1\right) /2$. \ Consequently,
all such relationships have now been found.

\section{APPENDIX E: \ Derivatives of Isotropic\ Tensors}

\subsection{Notation}

\qquad The objective of this appendix is to develop succinct notation for
isotropic tensors and their derivatives with specific attention to their
first-order divergence and their Laplacian. \ Those derivatives appear in (%
\ref{1iso}). \ A derivation of those derivatives operating on an isotropic
tensor that is symmetric under interchange of any pair of indexes is given.

\qquad First, notation is developed: $\ \mathbf{\delta }_{\left[ 2P\right] }$
is the product of $P$ Kronecker deltas having $2P$ distinct indexes. $\ $%
For\ example, $\mathbf{\delta }_{\left[ 6\right] }=\delta _{ij}\delta
_{kl}\delta _{mn}$, where $\delta _{ij}=1$ if $i=j$ and $\delta _{ij}=0$ if $%
i\neq j$. $\ \mathbf{W}_{\left[ N\right] }$ is the product of $N$ factors $%
\frac{r_{i}}{r}$ each with an index distinct from the other indexes. \ For
example, $\mathbf{W}_{\left[ 4\right] }=\frac{r_{i}}{r}\frac{r_{j}}{r}\frac{%
r_{k}}{r}\frac{r_{l}}{r}$. \ For convenience, define 
\begin{equation}
\mathbf{\delta }_{\left[ 0\right] }\equiv 1,\mathbf{\delta }_{\left[ -2%
\right] }\equiv 0,\mathbf{W}_{\left[ 0\right] }\equiv 1,\mathbf{W}_{\left[ -1%
\right] }\equiv 0,\mathbf{W}_{\left[ -2\right] }\equiv 0.  \label{convenient}
\end{equation}%
\ The plural of $\mathbf{\delta }$ is $\mathbf{\delta }$s and that of $%
\mathbf{W}$ is $\mathbf{W}$s. \ It is understood that products of $\mathbf{W}
$s (e.g., $\mathbf{W}_{\left[ N\right] }\mathbf{W}_{\left[ K\right] }$)\ and
of $\mathbf{\delta }$s\ (e.g., $\mathbf{\delta }_{\left[ 2N\right] }\mathbf{%
\delta }_{\left[ 2K\right] }$)\ and of $\mathbf{W}$s\ with $\mathbf{\delta }$%
s\ (e.g., $\mathbf{W}_{\left[ N\right] }\mathbf{\delta }_{\left[ 2K\right] }$%
)\ have all distinct indexes. \ Then, the $\mathbf{W}$s factor, e.g., $%
\mathbf{W}_{\left[ 4\right] }=\mathbf{W}_{\left[ 1\right] }\mathbf{W}_{\left[
3\right] }=\mathbf{W}_{\left[ 2\right] }\mathbf{W}_{\left[ 2\right] }=%
\mathbf{W}_{\left[ 1\right] }\mathbf{W}_{\left[ 1\right] }\mathbf{W}_{\left[
2\right] }$, etc., and the $\mathbf{\delta }$s likewise factor. \ The
operation \textquotedblleft contraction\textquotedblright\ means to set two
indexes equal and sum over their range of values; the summation convention
over repeated Roman indexes is used, e.g., $\delta _{ii}\equiv \overset{3}{%
\underset{i=1}{\sum }}\delta _{ii}=\delta _{11}+\delta _{22}+\delta _{33}=3$%
. \ A contraction of two indexes of $\mathbf{\delta }_{\left[ 2P\right] }$\
produces either $3\mathbf{\delta }_{\left[ 2\left( P-1\right) \right] }$ or $%
\mathbf{\delta }_{\left[ 2\left( P-1\right) \right] }$\ depending on whether
the two indexes are on the same Kronecker delta or different ones,
respectively. \ The contraction of $\mathbf{W}_{\left[ N\right] }$ on two
indexes produces $\mathbf{W}_{\left[ N-1\right] }$ because $\frac{r_{i}}{r}%
\frac{r_{i}}{r}=\frac{r^{2}}{r^{2}}=1$. \ Consider the contraction of $\frac{%
r_{i}}{r}$ with $\mathbf{W}_{\left[ N\right] }\mathbf{\delta }_{\left[ 2P%
\right] }$. \ If the index $i$ is in $\mathbf{W}_{\left[ N\right] }$ then
the contraction $\frac{r_{i}}{r}\mathbf{W}_{\left[ N\right] }\mathbf{\delta }%
_{\left[ 2P\right] }$\ is $\mathbf{W}_{\left[ N-1\right] }\mathbf{\delta }_{%
\left[ 2P\right] }$ because $\frac{r_{i}}{r}\frac{r_{i}}{r}=1$. \ If the
index $i$ is in $\mathbf{\delta }_{\left[ 2P\right] }$\ then the contraction 
$\frac{r_{i}}{r}\mathbf{W}_{\left[ N\right] }\mathbf{\delta }_{\left[ 2P%
\right] }$\ is $\mathbf{W}_{\left[ N+1\right] }\mathbf{\delta }_{\left[
2\left( P-1\right) \right] }$ because $\frac{r_{i}}{r}\delta _{ij}=\frac{%
r_{j}}{r}$. \ The notation $\left\Updownarrow \circ \right\Updownarrow
_{j}^{N}$means the sum of $N$\ terms where each term contains a distinct
index $j$, and the index $j$ is interchanged with all implied indexes, but $%
j $ is not interchanged with any explicit index. \ For example, 
\begin{eqnarray}
\left\Updownarrow \mathbf{W}_{\left[ 2\right] }\delta
_{ij}\right\Updownarrow _{j}^{3} &=&\frac{r_{k}}{r}\frac{r_{l}}{r}\delta
_{ij}+\frac{r_{k}}{r}\frac{r_{j}}{r}\delta _{il}+\frac{r_{j}}{r}\frac{r_{l}}{%
r}\delta _{ik},  \label{sigmaexample} \\
\text{ and }\left\Updownarrow \mathbf{W}_{\left[ 1\right] }\mathbf{\delta }_{%
\left[ 2\right] }\frac{r_{i}}{r}\frac{r_{j}}{r}\right\Updownarrow _{j}^{4}
&=&\frac{r_{k}}{r}\delta _{nm}\frac{r_{i}}{r}\frac{r_{j}}{r}+\frac{r_{j}}{r}%
\delta _{nm}\frac{r_{i}}{r}\frac{r_{k}}{r}+\frac{r_{k}}{r}\delta _{jm}\frac{%
r_{i}}{r}\frac{r_{n}}{r}+\frac{r_{k}}{r}\delta _{nj}\frac{r_{i}}{r}\frac{%
r_{m}}{r},  \notag
\end{eqnarray}%
where, $i$ is an explicit index and is therefore not interchanged with $j$.

\qquad French braces (i.e., $\left\{ \circ \right\} $)\ means: \ add all
such distinct terms required to make the tensor symmetric under interchange
of any pair of indexes. \ For example, 
\begin{equation*}
\left\{ \mathbf{W}_{\left[ 2\right] }\mathbf{\delta }_{\left[ 2\right]
}\right\} =\left\{ \frac{r_{i}}{r}\frac{r_{j}}{r}\delta _{kl}\right\} =\frac{%
r_{i}}{r}\frac{r_{j}}{r}\delta _{kl}+\frac{r_{i}}{r}\frac{r_{l}}{r}\delta
_{kj}+\frac{r_{l}}{r}\frac{r_{j}}{r}\delta _{ki}+\frac{r_{k}}{r}\frac{r_{l}}{%
r}\delta _{ij}+\frac{r_{k}}{r}\frac{r_{j}}{r}\delta _{il}+\frac{r_{i}}{r}%
\frac{r_{k}}{r}\delta _{jl}.
\end{equation*}
Note that terms that are necessarily equal do not appear; i.e., since $\frac{%
r_{i}}{r}\frac{r_{j}}{r}\delta _{kl}$\ appears, neither $\frac{r_{j}}{r}%
\frac{r_{i}}{r}\delta _{kl}$\ nor $\frac{r_{i}}{r}\frac{r_{j}}{r}\delta
_{lk} $\ appear. \ Because of the commutative law of addition, $\left\{
\circ \right\} $\ commutes with addition; e.g., $\left\{ \mathbf{W}_{\left[ N%
\right] }\right\} +\left\{ \mathbf{W}_{\left[ Q\right] }\mathbf{\delta }_{%
\left[ 2P\right] }\right\} =\left\{ \mathbf{W}_{\left[ N\right] }+\mathbf{W}%
_{\left[ Q\right] }\mathbf{\delta }_{\left[ 2P\right] }\right\} $. \ Because
of the distributive law of multiplication, multiplication by a scalar
function commutes with the $\left\{ \circ \right\} $\ notation; i.e., $%
A\left( r\right) \left\{ \mathbf{W}_{\left[ N\right] }\mathbf{\delta }_{%
\left[ 2P\right] }\right\} =\left\{ A\left( r\right) \mathbf{W}_{\left[ N%
\right] }\mathbf{\delta }_{\left[ 2P\right] }\right\} $.

\qquad The number of terms in various sums, $\left\{ \circ \right\} $, is
required repeatedly: $\ \left\{ \mathbf{\delta }_{\left[ 2P\right] }\right\} 
$ has $\left( 2P-1\right) !!=\left( 2P-1\right) \left( 2P-3\right) \left(
2P-5\right) \cdot \cdot \cdot \left( 1\right) $ terms. \ Since\ $\mathbf{W}_{%
\left[ N\right] }\mathbf{\delta }_{\left[ 2P\right] }$\ has $2P+N$\
indexes,\ $\left\{ \mathbf{W}_{\left[ N\right] }\mathbf{\delta }_{\left[ 2P%
\right] }\right\} $ has $\binom{2P+N}{N}\left( 2P-1\right) !!$ terms, where
the binomial coefficient $\binom{2P+N}{N}$\ is the number of ways of
selecting the $N$\ indexes in $\mathbf{W}_{\left[ N\right] }$\ from the
total $2P+N$ indexes. \ If $i$ is an index in $\left\{ \mathbf{W}_{\left[ N%
\right] }\mathbf{\delta }_{\left[ 2P\right] }\right\} $, then $\left\{ 
\mathbf{W}_{\left[ N\right] }\mathbf{\delta }_{\left[ 2P\right] }\right\} $
has $\binom{2P+N-1}{N}\left( 2P-1\right) !!$ terms in which $i$ appears in a
Kronecker delta because there are $N$\ indexes to select for $\mathbf{W}_{%
\left[ N\right] }$\ from the remaining $2P+N-1$ indexes. \ Similarly, $%
\left\{ \mathbf{W}_{\left[ N\right] }\mathbf{\delta }_{\left[ 2P\right]
}\right\} $ has $\binom{2P+N-1}{N-1}\left( 2P-1\right) !!$ terms\ in which $%
i $ appears in a factor $\left( r_{i}/r\right) $ because there are $N-1$\
indexes remaining to select for $\mathbf{W}_{\left[ N\right] }$\ from the
remaining $2P+N-1$ indexes. \ Note that $\left\{ \mathbf{W}_{\left[ N\right]
}\right\} $ has only 1 term. \ Hence, $\left\{ \mathbf{W}_{\left[ N\right]
}\right\} =\mathbf{W}_{\left[ N\right] }$. \ In summary, 
\begin{eqnarray}
&&\left\{ \mathbf{W}_{\left[ N\right] }\mathbf{\delta }_{\left[ 2P\right]
}\right\} \text{ has }\binom{2P+N}{N}\left( 2P-1\right) !!\text{ terms};
\label{numbterms} \\
&&\left\{ \mathbf{W}_{\left[ N\right] }\mathbf{\delta }_{\left[ 2P\right]
}\right\} \text{ has }\binom{2P+N-1}{N}\left( 2P-1\right) !!\text{ terms
with }i\text{\ in }\mathbf{\delta }_{\left[ 2P\right] };
\label{numbtermsdelta} \\
&&\left\{ \mathbf{W}_{\left[ N\right] }\mathbf{\delta }_{\left[ 2P\right]
}\right\} \text{ has }\binom{2P+N-1}{N-1}\left( 2P-1\right) !!\text{ terms
with }i\text{\ in }\mathbf{W}_{\left[ N\right] }.  \label{numbtermsiW}
\end{eqnarray}
The sum of the number of terms in (\ref{numbtermsdelta}-\ref{numbtermsiW}),
namely $\binom{2P+N-1}{N}\left( 2P-1\right) !!+\binom{2P+N-1}{N-1}\left(
2P-1\right) !!=\left( 2P+N-1\right) !\left[ \frac{2P}{N!\left( 2P\right) !}+%
\frac{N}{N!\left( 2P\right) !}\right] \left( 2P-1\right) !!=\binom{2P+N}{N}%
\left( 2P-1\right) !!$, agrees with the total number of terms in (\ref%
{numbterms}).

\qquad Now, rules for differentiation of symmetric, isotropic tensors are
developed. \ Note the identity 
\begin{equation}
\partial _{r_{i}}\left( r_{j}/r\right) =\left[ \delta _{ij}-\left(
r_{i}r_{j}/r^{2}\right) \right] /r=\left( \mathbf{\delta }_{\left[ 2\right]
}-\mathbf{W}_{\left[ 2\right] }\right) /r,  \label{gradW1}
\end{equation}
from which it follows that 
\begin{equation}
\partial _{r_{i}}\left( r_{i}/r\right) =\left( 3-1\right) /r=2/r\text{, and }%
r_{i}\partial _{r_{i}}\left( r_{j}/r\right) =\left[ r_{j}-\left(
r^{2}r_{j}/r^{2}\right) \right] /r=0.  \label{helps}
\end{equation}
The latter formula greatly simplifies the divergence of $\mathbf{W}_{\left[ N%
\right] }$ because $\partial _{r_{i}}$ operating on $\mathbf{W}_{\left[ N%
\right] }\ $vanishes\ when it operates on any factor other than the factor $%
\frac{r_{i}}{r}$\ within $\mathbf{W}_{\left[ N\right] }$.

\subsection{Gradient and Divergence}

\qquad The divergence and gradient of $\mathbf{W}_{\left[ N\right] }$\ are
needed. If $i$\ is an index in $\mathbf{W}_{\left[ N\right] }$\ then the
divergence of $\mathbf{W}_{\left[ N\right] }$\ is denoted by $\nabla _{%
\mathbf{r}}\cdot \mathbf{W}_{\left[ N\right] }=\partial _{r_{i}}\mathbf{W}_{%
\left[ N\right] }$. \ Application of (\ref{helps}) gives $\partial _{r_{i}}%
\mathbf{W}_{\left[ 1\right] }=\partial _{r_{i}}\left( r_{i}/r\right) =2/r$,
and $\partial _{r_{i}}\mathbf{W}_{\left[ N\right] }=\mathbf{W}_{\left[ N-1%
\right] }\partial _{r_{i}}\left( r_{i}/r\right) =\frac{2}{r}\mathbf{W}_{%
\left[ N-1\right] }$. \ The gradient of $\mathbf{W}_{\left[ N\right] }$ is
denoted by $\partial _{r_{i}}\mathbf{W}_{\left[ N\right] }$\ where $i$ is
not an index in $\mathbf{W}_{\left[ N\right] }$. \ From the differentiation
chain rule, $\partial _{r_{i}}\mathbf{W}_{\left[ N\right] }$ is the sum of $%
N $ terms, each of which has the form $\mathbf{W}_{\left[ N-1\right]
}\partial _{r_{i}}\left( r_{j}/r\right) $. \ Therefore, by use of (\ref%
{gradW1}), $\partial _{r_{i}}\mathbf{W}_{\left[ N\right] }=\left\Updownarrow 
\mathbf{W}_{\left[ N-1\right] }\left[ \delta _{ij}-\left(
r_{i}r_{j}/r^{2}\right) \right] /r\right\Updownarrow _{j}^{N}=\frac{1}{r}%
\left\Updownarrow \mathbf{W}_{\left[ N-1\right] }\delta
_{ij}\right\Updownarrow _{j}^{N}-\frac{N}{r}\mathbf{W}_{\left[ N+1\right] }$%
, where use was made of $\left\Updownarrow \mathbf{W}_{\left[ N-1\right] }%
\left[ -\left( r_{i}r_{j}/r^{2}\right) \right] /r\right\Updownarrow _{j}^{N}=%
\left[ -\left( r_{i}/r\right) /r\right] \left\Updownarrow \mathbf{W}_{\left[
N-1\right] }\left( r_{j}/r\right) \right\Updownarrow _{j}^{N}=-\left(
r_{i}/r\right) /r\left( N\mathbf{W}_{\left[ N\right] }\right) =-\frac{N}{r}%
\mathbf{W}_{\left[ N+1\right] }$, because $\left\Updownarrow \mathbf{W}_{%
\left[ N-1\right] }\left( r_{j}/r\right) \right\Updownarrow _{j}^{N}$\ is
the sum of $N$ identical terms each equal to $\mathbf{W}_{\left[ N\right] }$%
. \ In summary, 
\begin{eqnarray}
\text{If }i\text{ is in }\mathbf{W}_{\left[ N\right] }\text{\ then }\nabla _{%
\mathbf{r}}\cdot \mathbf{W}_{\left[ N\right] } &=&\partial _{r_{i}}\mathbf{W}%
_{\left[ N\right] }=\frac{2}{r}\mathbf{W}_{\left[ N-1\right] }.
\label{gdiverge} \\
\text{If }i\text{ is not in }\mathbf{W}_{\left[ N\right] }\text{\ then }%
\partial _{r_{i}}\mathbf{W}_{\left[ N\right] } &=&\frac{1}{r}%
\left\Updownarrow \mathbf{W}_{\left[ N-1\right] }\delta
_{ij}\right\Updownarrow _{j}^{N}-\frac{N}{r}\mathbf{W}_{\left[ N+1\right] },
\label{gradW}
\end{eqnarray}

\qquad Consider the divergence of $\mathbf{W}_{\left[ N\right] }\mathbf{%
\delta }_{\left[ 2P\right] }.$ \ If the index $i$ is in $\mathbf{W}_{\left[ N%
\right] }$ then, from (\ref{gdiverge}), $\partial _{r_{i}}\left( \mathbf{W}_{%
\left[ N\right] }\mathbf{\delta }_{\left[ 2P\right] }\right) =\mathbf{\delta 
}_{\left[ 2P\right] }\mathbf{W}_{\left[ N\mid 1\right] }=\frac{2}{r}\mathbf{%
\delta }_{\left[ 2P\right] }\mathbf{W}_{\left[ N-1\right] }$. \ If the index 
$i$ is in $\mathbf{\delta }_{\left[ 2P\right] }$, then (given that index $k$%
\ is not in $\mathbf{W}_{\left[ N\right] }$) $\partial _{r_{i}}\left( 
\mathbf{W}_{\left[ N\right] }\mathbf{\delta }_{\left[ 2P\right] }\right) =%
\mathbf{\delta }_{\left[ 2\left( P-1\right) \right] }\delta _{ik}\partial
_{r_{i}}\mathbf{W}_{\left[ N\right] }=\mathbf{\delta }_{\left[ 2\left(
P-1\right) \right] }\partial _{r_{k}}\mathbf{W}_{\left[ N\right] }=\mathbf{%
\delta }_{\left[ 2\left( P-1\right) \right] }\frac{1}{r}\left\Updownarrow 
\mathbf{W}_{\left[ N-1\right] }\delta _{kj}\right\Updownarrow _{j}^{N}-%
\mathbf{\delta }_{\left[ 2\left( P-1\right) \right] }\frac{N}{r}\mathbf{W}_{%
\left[ N+1\right] }$, where the last expression follows from (\ref{gradW}).
\ In summary, 
\begin{eqnarray}
&&\text{If }i\text{ is in }\mathbf{W}_{\left[ N\right] }\text{ then }  \notag
\\
\partial _{r_{i}}\left( \mathbf{W}_{\left[ N\right] }\mathbf{\delta }_{\left[
2P\right] }\right) &=&\frac{2}{r}\mathbf{\delta }_{\left[ 2P\right] }\mathbf{%
W}_{\left[ N-1\right] },  \label{div-i-in-W1} \\
&&\text{If }i\text{ is in }\mathbf{\delta }_{\left[ 2P\right] }\text{ then }
\notag \\
\partial _{r_{i}}\left( \mathbf{W}_{\left[ N\right] }\mathbf{\delta }_{\left[
2P\right] }\right) &=&\mathbf{\delta }_{\left[ 2\left( P-1\right) \right] }%
\frac{1}{r}\left\Updownarrow \mathbf{W}_{\left[ N-1\right] }\delta
_{kj}\right\Updownarrow _{j}^{N}-\mathbf{\delta }_{\left[ 2\left( P-1\right) %
\right] }\frac{N}{r}\mathbf{W}_{\left[ N+1\right] }.  \label{div-i-in-delta}
\end{eqnarray}

\qquad The above results allow evaluation of the divergence $\partial
_{r_{i}}\left\{ \mathbf{W}_{\left[ N\right] }\mathbf{\delta }_{\left[ 2P%
\right] }\right\} \equiv \nabla _{\mathbf{r}}\cdot \left\{ \mathbf{W}_{\left[
N\right] }\mathbf{\delta }_{\left[ 2P\right] }\right\} $. \ It follows from
use of (\ref{div-i-in-W1}) and the distributive law of multiplication and
the fact that the number of terms in $\left\{ \mathbf{W}_{\left[ N\right] }%
\mathbf{\delta }_{\left[ 2P\right] }\right\} $\ in which $i$ appears in the
factor $\frac{r_{i}}{r}$ is the same as the number of terms in $\left\{ 
\mathbf{W}_{\left[ N-1\right] }\mathbf{\delta }_{\left[ 2P\right] }\right\} $
[see (\ref{numbterms}-\ref{numbtermsiW})], that for those terms in which $i$
is in $\mathbf{W}_{\left[ N\right] }$, the divergence of $\left\{ \mathbf{W}%
_{\left[ N\right] }\mathbf{\delta }_{\left[ 2P\right] }\right\} $ yields $%
\frac{2}{r}\left\{ \mathbf{W}_{\left[ N-1\right] }\mathbf{\delta }_{\left[ 2P%
\right] }\right\} $. \ Similar use of (\ref{div-i-in-delta}) gives that for
those terms in which $i$ is in $\mathbf{\delta }_{\left[ 2P\right] }$, the
divergence of $\left\{ \mathbf{W}_{\left[ N\right] }\mathbf{\delta }_{\left[
2P\right] }\right\} $ yields $\frac{2P}{r}\left\{ \mathbf{W}_{\left[ N-1%
\right] }\mathbf{\delta }_{\left[ 2P\right] }\right\} -\frac{N\left(
N+1\right) }{r}\left\{ \mathbf{W}_{\left[ N+1\right] }\mathbf{\delta }_{%
\left[ 2\left( P-1\right) \right] }\right\} $. \ Thus, 
\begin{eqnarray}
&&\nabla _{\mathbf{r}}\cdot \left\{ \mathbf{W}_{\left[ N\right] }\mathbf{%
\delta }_{\left[ 2P\right] }\right\}  \notag \\
&=&\frac{2}{r}\left\{ \mathbf{W}_{\left[ N-1\right] }\mathbf{\delta }_{\left[
2P\right] }\right\} +\left[ \frac{2P}{r}\left\{ \mathbf{W}_{\left[ N-1\right]
}\mathbf{\delta }_{\left[ 2P\right] }\right\} -\frac{N\left( N+1\right) }{r}%
\left\{ \mathbf{W}_{\left[ N+1\right] }\mathbf{\delta }_{\left[ 2\left(
P-1\right) \right] }\right\} \right]  \notag \\
&=&\frac{2}{r}\left( P+1\right) \left\{ \mathbf{W}_{\left[ N-1\right] }%
\mathbf{\delta }_{\left[ 2P\right] }\right\} -\frac{N\left( N+1\right) }{r}%
\left\{ \mathbf{W}_{\left[ N+1\right] }\mathbf{\delta }_{\left[ 2\left(
P-1\right) \right] }\right\} .  \label{symdiverg}
\end{eqnarray}%
\ Because of the definitions in (\ref{convenient}), (\ref{symdiverg})
remains valid if $N$ is $0$ or $1$ or if $P$ is $0$ or $1$.

\qquad Derivation of the formula for the divergence of an isotropic tensor
requires evaluation of the contraction $\left\{ \mathbf{W}_{\left[ N\right] }%
\mathbf{\delta }_{\left[ 2P\right] }\right\} \frac{r_{i}}{r}$. \ From (\ref%
{numbtermsdelta}), in $\left\{ \mathbf{W}_{\left[ N\right] }\mathbf{\delta }%
_{\left[ 2P\right] }\right\} $\ there are $\binom{2P+N-1}{N}\left(
2P-1\right) !!$\ occurrences of the index $i$ within $\mathbf{\delta }_{%
\left[ 2P\right] }$ and each gives the contraction $\delta _{ij}\frac{r_{i}}{%
r}=\frac{r_{j}}{r}$, which decreases $P$ by unity and increases $N$ by unity
thereby producing several $\left\{ \mathbf{W}_{\left[ N+1\right] }\mathbf{%
\delta }_{\left[ 2\left( P-1\right) \right] }\right\} $. \ From (\ref%
{numbterms}), there are $\binom{2\left( P-1\right) +\left( N+1\right) }{%
\left( N+1\right) }\left( 2\left( P-1\right) -1\right) !!$\ terms in a $%
\left\{ \mathbf{W}_{\left[ N+1\right] }\mathbf{\delta }_{\left[ 2\left(
P-1\right) \right] }\right\} $; thus the number of $\left\{ \mathbf{W}_{%
\left[ N+1\right] }\mathbf{\delta }_{\left[ 2\left( P-1\right) \right]
}\right\} $ so produced is 
\begin{equation*}
\left[ \binom{2P+N-1}{N}\left( 2P-1\right) !!\right] /\left[ \binom{2\left(
P-1\right) +\left( N+1\right) }{\left( N+1\right) }\left( 2\left( P-1\right)
-1\right) !!\right] =\left( N+1\right) .
\end{equation*}%
\ From (\ref{numbtermsiW}), the contraction $\left\{ \mathbf{W}_{\left[ N%
\right] }\mathbf{\delta }_{\left[ 2P\right] }\right\} \frac{r_{i}}{r}$
contains $\binom{2P+N-1}{N-1}\left( 2P-1\right) !!$ terms\ in which $i$
appears within $\mathbf{W}_{\left[ N\right] }$ and each results in the
contraction $\frac{r_{i}}{r}\frac{r_{i}}{r}=1$, which decreases $N$ by
unity. \ The number of $\left\{ \mathbf{W}_{\left[ N-1\right] }\mathbf{%
\delta }_{\left[ 2P\right] }\right\} $\ so produced is $\left[ \binom{2P+N-1%
}{N-1}\left( 2P-1\right) !!\right] /\left[ \binom{2P+\left( N-1\right) }{%
\left( N-1\right) }\left( 2P-1\right) !!\right] =1$ because $\left\{ \mathbf{%
W}_{\left[ N-1\right] }\mathbf{\delta }_{\left[ 2P\right] }\right\} $\ has $%
\binom{2P+\left( N-1\right) }{\left( N-1\right) }\left( 2P-1\right) !!$\
terms, which is also the number of terms given in (\ref{numbtermsiW}). \
Thus, 
\begin{equation}
\text{contraction on }i\text{: }\left\{ \mathbf{W}_{\left[ N\right] }\mathbf{%
\delta }_{\left[ 2P\right] }\right\} \frac{r_{i}}{r}=\left( N+1\right)
\left\{ \mathbf{W}_{\left[ N+1\right] }\mathbf{\delta }_{\left[ 2\left(
P-1\right) \right] }\right\} +\left\{ \mathbf{W}_{\left[ N-1\right] }\mathbf{%
\delta }_{\left[ 2P\right] }\right\} .  \label{symcontract}
\end{equation}

\qquad The general isotropic formula for a tensor $\mathbf{A}_{\left[ N%
\right] }\left( \mathbf{r}\right) $ of order $N$ that is symmetric under
interchange of any pair of indexes is 
\begin{eqnarray}
\mathbf{A}_{\left[ N\right] }\left( \mathbf{r}\right) &=&A_{0}\left(
r\right) \left\{ \mathbf{W}_{\left[ N\right] }\right\} +A_{1}\left( r\right)
\left\{ \mathbf{W}_{\left[ N-2\right] }\mathbf{\delta }_{\left[ 2\right]
}\right\} +A_{2}\left( r\right) \left\{ \mathbf{W}_{\left[ N-4\right] }%
\mathbf{\delta }_{\left[ 4\right] }\right\} +\cdot \cdot \cdot +T_{last} 
\notag \\
&=&\overset{M}{\underset{P=0}{\sum }}A_{P}\left( r\right) \left\{ \mathbf{W}%
_{\left[ N-2P\right] }\mathbf{\delta }_{\left[ 2P\right] }\right\} ,
\label{isoformula}
\end{eqnarray}
where the $A_{0}\left( r\right) $, $A_{1}\left( r\right) $, etc., are scalar
functions of $r$, and $T_{last}$ is the last term. \ Note that for brevity
in this appendix, the subscript $N$ has been omitted from $A_{N,0}\left(
r\right) $, $A_{N,1}\left( r\right) $, etc. \ If $N$ is even, then $%
T_{last}=A_{N/2}\left( r\right) \left\{ \mathbf{\delta }_{\left[ N\right]
}\right\} $ and $M=N/2$. \ If $N$ is odd, then $T_{last}=A_{\left(
N-1\right) /2}\left( r\right) \left\{ \mathbf{W}_{\left[ 1\right] }\mathbf{%
\delta }_{\left[ N-1\right] }\right\} $ and $M=\left( N-1\right) /2$.

\qquad All of the foregoing has set the stage for efficient derivation of a
formula for the divergence $\nabla _{\mathbf{r}}\cdot \mathbf{A}_{\left[ N%
\right] }\left( \mathbf{r}\right) $. \ Also needed is the fact that the
gradient of a scalar function of $r\equiv \sqrt{r_{i}r_{i}}$ is $\partial
_{r_{i}}A\left( r\right) =\left( \partial _{r_{i}}r\right) \partial
_{r}A\left( r\right) =\frac{r_{i}}{r}\partial _{r}A\left( r\right) =\mathbf{W%
}_{\left[ 1\right] }\partial _{r}A\left( r\right) $. \ Consider the
divergence of a term in (\ref{isoformula}). \ By use of the differentiation
chain rule, and substitution of (\ref{symcontract}) and (\ref{symdiverg}),%
\begin{equation*}
\nabla _{\mathbf{r}}\cdot \left[ A_{P}\left( r\right) \left\{ \mathbf{W}_{%
\left[ N-2P\right] }\mathbf{\delta }_{\left[ 2P\right] }\right\} \right] =
\end{equation*}
\begin{eqnarray}
&=&\left\{ \mathbf{W}_{\left[ N-2P\right] }\mathbf{\delta }_{\left[ 2P\right]
}\right\} \frac{r_{i}}{r}\partial _{r}A_{P}\left( r\right) +A_{P}\left(
r\right) \nabla _{\mathbf{r}}\cdot \left\{ \mathbf{W}_{\left[ N-2P\right] }%
\mathbf{\delta }_{\left[ 2P\right] }\right\}  \notag \\
&=&\left[ \left( N-2P+1\right) \left\{ \mathbf{W}_{\left[ N-2P+1\right] }%
\mathbf{\delta }_{\left[ 2\left( P-1\right) \right] }\right\} +\left\{ 
\mathbf{W}_{\left[ N-2P-1\right] }\mathbf{\delta }_{\left[ 2P\right]
}\right\} \right] \partial _{r}A_{P}\left( r\right)  \notag \\
&&+A_{P}\left( r\right) \left[ \frac{2}{r}\left( P+1\right) \left\{ \mathbf{W%
}_{\left[ N-2P-1\right] }\mathbf{\delta }_{\left[ 2P\right] }\right\} \right]
\notag \\
&&-A_{P}\left( r\right) \left[ \frac{\left( N-2P\right) \left( N-2P+1\right) 
}{r}\left\{ \mathbf{W}_{\left[ N-2P+1\right] }\mathbf{\delta }_{\left[
2\left( P-1\right) \right] }\right\} \right]  \notag \\
&=&B_{N,P}\left( r\right) \left\{ \mathbf{W}_{\left[ N-2P+1\right] }\mathbf{%
\delta }_{\left[ 2\left( P-1\right) \right] }\right\} +C_{P}\left( r\right)
\left\{ \mathbf{W}_{\left[ N-2P-1\right] }\mathbf{\delta }_{\left[ 2P\right]
}\right\} ,  \label{divoneterm}
\end{eqnarray}%
where $B_{N,P}\left( r\right) $ and $C_{P}\left( r\right) $ are defined by
the following operators, $O_{B}\left( N,P\right) $\ and $O_{C}\left(
P\right) $, operating on $A_{P}\left( r\right) $: 
\begin{eqnarray}
B_{N,P}\left( r\right) &\equiv &O_{B}\left( N,P\right) A_{P}\left( r\right) ,%
\text{ where }  \label{1operator} \\
O_{B}\left( N,P\right) &\equiv &\left( N-2P+1\right) \left[ \partial _{r}-%
\frac{N-2P}{r}\right] ,  \notag \\
C_{P}\left( r\right) &\equiv &O_{C}\left( P\right) A_{P}\left( r\right) 
\text{ where }O_{C}\left( P\right) \equiv \left[ \partial _{r}+\frac{2}{r}%
\left( P+1\right) \right] .  \label{2operator}
\end{eqnarray}%
Thereby, the divergence of (\ref{isoformula}) is 
\begin{equation}
\nabla _{\mathbf{r}}\cdot \mathbf{A}_{\left[ N\right] }\left( \mathbf{r}%
\right) =\overset{M}{\underset{P=0}{\sum }}B_{N,P}\left( r\right) \left\{ 
\mathbf{W}_{\left[ N-2P+1\right] }\mathbf{\delta }_{\left[ 2\left(
P-1\right) \right] }\right\} +\overset{M}{\underset{P=0}{\sum }}C_{P}\left(
r\right) \left\{ \mathbf{W}_{\left[ N-2P-1\right] }\mathbf{\delta }_{\left[
2P\right] }\right\} ,  \label{1divDn}
\end{equation}%
where (\ref{divoneterm}) and (\ref{convenient}) were used.

\qquad Now, (\ref{1divDn}) can be checked by comparison with the divergence
performed on the explicit-index formulas for symmetric, isotropic tensors of
rank 1 to 4. \ The lowest-order tensor for which the divergence is defined
is a vector (i.e., $N=1)$, in which case (\ref{1divDn}) gives 
\begin{equation*}
\partial _{r_{i}}A_{i}\left( \mathbf{r}\right) =\partial _{r}A_{0}\left(
r\right) +\frac{2}{r}A_{0}\left( r\right) ,
\end{equation*}%
which is easily verified by evaluating the divergence of a isotropic vector,
namely $\partial _{r_{i}}\left[ A_{0}\left( r\right) \frac{r_{i}}{r}\right] $%
. \ Expressed with explicit indexes as well as in the implicit-index form of
(\ref{isoformula}), isotropic tensors of rank 2 to 4 that are symmetric
under interchange of any pair of indexes are: 
\begin{equation}
A_{ij}\left( \mathbf{r}\right) =A_{0}\left( r\right) \frac{r_{i}}{r}\frac{%
r_{j}}{r}+A_{1}\left( r\right) \delta _{ij}=A_{0}\left( r\right) \mathbf{W}_{%
\left[ 2\right] }+A_{1}\left( r\right) \mathbf{\delta }_{\left[ 2\right] }.
\label{isoexample}
\end{equation}%
\begin{equation}
A_{ijk}\left( \mathbf{r}\right) =A_{0}\left( r\right) \frac{r_{i}}{r}\frac{%
r_{j}}{r}\frac{r_{k}}{r}+A_{1}\left( r\right) \left( \frac{r_{i}}{r}\delta
_{jk}+\frac{r_{j}}{r}\delta _{ik}+\frac{r_{k}}{r}\delta _{ij}\right)
=A_{0}\left( r\right) \mathbf{W}_{\left[ 3\right] }+A_{1}\left( r\right)
\left\{ \mathbf{W}_{\left[ 1\right] }\mathbf{\delta }_{\left[ 2\right]
}\right\} .  \label{isoexample2}
\end{equation}%
\begin{eqnarray}
A_{ijkl}\left( \mathbf{r}\right) &=&A_{0}\left( r\right) \frac{%
r_{i}r_{j}r_{k}r_{l}}{r^{4}}+A_{1}\left( r\right) \left( \frac{r_{i}r_{j}}{%
r^{2}}\delta _{kl}+\frac{r_{i}r_{k}}{r^{2}}\delta _{jl}+\frac{r_{j}r_{k}}{%
r^{2}}\delta _{il}+\frac{r_{i}r_{l}}{r^{2}}\delta _{jk}+\frac{r_{j}r_{l}}{%
r^{2}}\delta _{ik}+\frac{r_{k}r_{l}}{r^{2}}\delta _{ij}\right)  \notag \\
&&+A_{2}\left( r\right) \left( \delta _{ij}\delta _{kl}+\delta _{ik}\delta
_{jl}+\delta _{jk}\delta _{il}\right)  \notag \\
&=&A_{0}\left( r\right) \mathbf{W}_{\left[ 4\right] }+A_{1}\left( r\right)
\left\{ \mathbf{W}_{\left[ 2\right] }\mathbf{\delta }_{\left[ 2\right]
}\right\} +A_{2}\left( r\right) \left\{ \mathbf{\delta }_{\left[ 4\right]
}\right\} .  \label{isoexample3}
\end{eqnarray}%
One can see the brevity of the implicit-index formula as the rank of the
tensor increases. \ The first-order divergences obtained by differentiating
the above explicit-index formulas as well as from (\ref{1divDn}) are: 
\begin{eqnarray}
\nabla _{\mathbf{r}}\cdot \mathbf{A}_{\left[ 2\right] }\left( \mathbf{r}%
\right) &=&\left[ \left( \partial _{r}+\frac{2}{r}\right) A_{0}\left(
r\right) +\partial _{r}A_{1}\left( r\right) \right] \frac{r_{j}}{r}
\label{1check} \\
&=&\left[ B_{2,1}\left( r\right) +C_{0}\left( r\right) \right] \left\{ 
\mathbf{W}_{\left[ 1\right] }\mathbf{\delta }_{\left[ 0\right] }\right\} .
\label{1checkimp}
\end{eqnarray}%
\begin{equation*}
\nabla _{\mathbf{r}}\cdot \mathbf{A}_{\left[ 3\right] }\left( \mathbf{r}%
\right) =\left[ \left( \partial _{r}+\frac{2}{r}\right) A_{0}\left( r\right)
+\left( 2\partial _{r}-\frac{2}{r}\right) A_{1}\left( r\right) \right] \frac{%
r_{j}r_{k}}{r^{2}}
\end{equation*}%
\begin{eqnarray}
&&+\left[ \left( \partial _{r}+\frac{4}{r}\right) A_{1}\left( r\right) %
\right] \delta _{jk}  \label{1check3rd} \\
&=&\left[ B_{3,1}\left( r\right) +C_{0}\left( r\right) \right] \left\{ 
\mathbf{W}_{\left[ 2\right] }\mathbf{\delta }_{\left[ 0\right] }\right\}
+C_{1}\left( r\right) \left\{ \mathbf{W}_{\left[ 0\right] }\mathbf{\delta }_{%
\left[ 2\right] }\right\} .  \label{1check3rdimp}
\end{eqnarray}

\begin{equation*}
\nabla _{\mathbf{r}}\cdot \mathbf{A}_{\left[ 4\right] }\left( \mathbf{r}%
\right) =
\end{equation*}%
\begin{eqnarray}
&=&\left[ \left( \partial _{r}+\frac{2}{r}\right) A_{0}\left( r\right)
+\left( 3\partial _{r}-\frac{6}{r}\right) A_{1}\left( r\right) \right] \frac{%
r_{j}r_{k}r_{l}}{r^{3}}  \notag \\
&&+\left[ \left( \partial _{r}+\frac{4}{r}\right) A_{1}\left( r\right)
+\partial _{r}A_{2}\left( r\right) \right] \left( \frac{r_{i}}{r}\delta
_{jk}+\frac{r_{j}}{r}\delta _{ik}+\frac{r_{k}}{r}\delta _{ij}\right)
\label{1check4th} \\
&=&\left[ B_{4,1}\left( r\right) +C_{0}\left( r\right) \right] \left\{ 
\mathbf{W}_{\left[ 3\right] }\mathbf{\delta }_{\left[ 0\right] }\right\}
+\left( B_{4,2}\left( r\right) +C_{1}\left( r\right) \right) \left\{ \mathbf{%
W}_{\left[ 1\right] }\mathbf{\delta }_{\left[ 2\right] }\right\} .
\label{1check4thimp}
\end{eqnarray}%
In the implicit-index formulas in (\ref{1checkimp}, \ref{1check3rdimp}, \ref%
{1check4thimp}), terms from (\ref{1divDn}) that are zero because of (\ref%
{convenient}) have been omitted. Equation (\ref{1divDn}) has been checked by
using the implicit-index formulas in (\ref{1checkimp}, \ref{1check3rdimp}, %
\ref{1check4thimp}) to obtain the explicit-index formulas in (\ref{1check}, %
\ref{1check3rd}, \ref{1check4th}), respectively.

\subsection{LaPlacian}

\qquad The Laplacian of a symmetric, isotropic tensor\ is also needed for
the term $2\nu \nabla _{\mathbf{r}}^{2}\mathbf{D}_{\left[ N\right] }$\ in (%
\ref{1iso}). \ 
\begin{eqnarray*}
\text{If }N-2P &\leq &0\text{\ \ then \ }\nabla ^{2}\mathbf{W}_{\left[ N-2P%
\right] }=0 \\
\text{If }N-2P &=&1\text{\ \ then \ }\nabla ^{2}\mathbf{W}_{\left[ 1\right]
}=\left\{ \partial _{r_{n}}\partial _{r_{n}}\frac{r_{j}}{r}\right\} \\
&=&\frac{-2}{r^{2}}\frac{r_{j}}{r}=\mathbf{W}_{\left[ 0\right] }\frac{-2}{%
r^{2}}\frac{r_{j}}{r}
\end{eqnarray*}%
Application of (\ref{1regslogident})\ to $\nabla ^{2}\mathbf{W}_{\left[ N-2P%
\right] }$ and use of (\ref{gradW1})-(\ref{helps}) gives, if $N-2P\geq 2$\
then, 
\begin{equation}
\nabla ^{2}\mathbf{W}_{\left[ N-2P\right] }=\left\{ \mathbf{W}_{\left[ N-2P-1%
\right] }\partial _{r_{n}}\partial _{r_{n}}\frac{r_{j}}{r}\right\} +2\left\{ 
\mathbf{W}_{\left[ N-2P-2\right] }\left( \partial _{r_{n}}\frac{r_{k}}{r}%
\right) \left( \partial _{r_{n}}\frac{r_{j}}{r}\right) \right\}  \label{LapW}
\end{equation}%
This also applies to $N-2P=1$\ \ if we let $\mathbf{W}_{\left[ -1\right] }=0$%
, and it applies to any $N-2P\leq 0$ if $\mathbf{W}_{\left[ N-2P\right] }=0$%
. \ Those conditions are a restatement of (\ref{convenient}) which is%
\begin{equation}
\mathbf{\delta }_{\left[ 0\right] }\equiv 1,\mathbf{\delta }_{\left[ -2%
\right] }\equiv 0,\mathbf{W}_{\left[ 0\right] }\equiv 1,\mathbf{W}_{\left[ -1%
\right] }\equiv 0,\mathbf{W}_{\left[ -2\right] }\equiv 0.
\end{equation}

\qquad Now, $\left\{ \mathbf{W}_{\left[ N-2P-1\right] }\partial
_{r_{n}}\partial _{r_{n}}\frac{r_{j}}{r}\right\} $ is $\binom{N-2P}{1}$
terms, each one is of the form $\mathbf{W}_{\left[ N-2P-1\right] }\partial
_{r_{n}}\partial _{r_{n}}\frac{r_{j}}{r}=\mathbf{W}_{\left[ N-2P-1\right]
}\left( \frac{-2}{r^{2}}\frac{r_{j}}{r}\right) =-\frac{2}{r^{2}}\mathbf{W}_{%
\left[ N-2P\right] }$. \ 

Also, $\left\{ \mathbf{W}_{\left[ N-2P-2\right] }\left( \partial _{r_{n}}%
\frac{r_{k}}{r}\right) \left( \partial _{r_{n}}\frac{r_{j}}{r}\right)
\right\} $ is $\binom{N-2P}{2}$ terms, each one is of the form

$\mathbf{W}_{\left[ N-2P-2\right] }\left( \partial _{r_{n}}\frac{r_{k}}{r}%
\right) \left( \partial _{r_{n}}\frac{r_{j}}{r}\right) =\mathbf{W}_{\left[
N-2P-2\right] }\frac{1}{r^{2}}\left( \delta _{kj}-\frac{r_{k}r_{j}}{r^{2}}%
\right) =\frac{1}{r^{2}}\mathbf{W}_{\left[ N-2P-2\right] }\delta _{ij}-\frac{%
1}{r^{2}}\mathbf{W}_{\left[ N-2P\right] }$. \ Thus, (\ref{LapW}) is \ 
\begin{equation}
\nabla ^{2}\mathbf{W}_{\left[ N-2P\right] }=-\frac{2}{r^{2}}\binom{N-2P}{1}%
\mathbf{W}_{\left[ N-2P\right] }+2\binom{N-2P}{2}\left( \frac{1}{r^{2}}%
\mathbf{W}_{\left[ N-2P-2\right] }\delta _{\left[ 2\right] }-\frac{1}{r^{2}}%
\mathbf{W}_{\left[ N-2P\right] }\right)  \label{LapW2}
\end{equation}%
The binomial coefficients\ prevent a nonzero term in (\ref{LapW2}) when $%
\mathbf{W}_{\left[ N-2P-1\right] }$\ or $\mathbf{W}_{\left[ N-2P-2\right] }$%
\ vanish in (\ref{LapW}) as required by definition (\ref{convenient})
provided \ that we define 
\begin{equation}
\binom{N-2P}{1}\equiv 0\text{ if }N-2P<1,\text{\ and }\binom{N-2P}{2}\equiv
0\ \text{ \ if }N-2P<2.  \label{binomialrule}
\end{equation}%
Of course, (\ref{binomialrule}) is consistent with $1/K!=0$ for $K<0$
(Abramowitz and Stegun, 1964, equation 6.1.7). \ Given (\ref{binomialrule}),
we can define, for brevity 
\begin{equation}
S_{N-2P}\equiv 2\binom{N-2P}{2}+2\binom{N-2P}{1}.  \label{Sdefinition}
\end{equation}%
Now (\ref{LapW2}) and (\ref{Sdefinition}) give 
\begin{equation}
\nabla ^{2}\mathbf{W}_{\left[ N-2P\right] }=\frac{2}{r^{2}}\binom{N-2P}{2}%
\mathbf{W}_{\left[ N-2P-2\right] }\delta _{\left[ 2\right] }-\frac{S_{N-2P}}{%
r^{2}}\mathbf{W}_{\left[ N-2P\right] }.  \label{Lapnexttolast}
\end{equation}%
Now, $\nabla ^{2}$ commutes with $\left\{ \text{\textbf{\ \ \ }}\right\} $\
because $\left\{ \text{\textbf{\ \ \ }}\right\} $\ \ is just a sum of
distinct terms, thus use of (\ref{Lapnexttolast}) gives

\begin{equation*}
\nabla ^{2}\left\{ \mathbf{W}_{\left[ N-2P\right] }\mathbf{\delta }_{\left[
2P\right] }\right\} =
\end{equation*}

\begin{eqnarray}
&=&\left\{ \left[ \nabla ^{2}\mathbf{W}_{\left[ N-2P\right] }\right] \mathbf{%
\delta }_{\left[ 2P\right] }\right\}  \notag \\
&=&\left\{ \left[ \frac{2}{r^{2}}\binom{N-2P}{2}\mathbf{W}_{\left[ N-2P-2%
\right] }\delta _{\left[ 2\right] }-\frac{S_{N-2P}}{r^{2}}\mathbf{W}_{\left[
N-2P\right] }\right] \mathbf{\delta }_{\left[ 2P\right] }\right\}  \notag \\
&=&\frac{2}{r^{2}}\left\{ \binom{N-2P}{2}\mathbf{W}_{\left[ N-2\left(
P+1\right) \right] }\mathbf{\delta }_{\left[ 2\left( P+1\right) \right] }-%
\frac{S_{N-2P}}{2}\mathbf{W}_{\left[ N-2P\right] }\mathbf{\delta }_{\left[ 2P%
\right] }\right\}  \label{LzeroV} \\
&=&\frac{R\left( N,P\right) }{r^{2}}\left\{ \mathbf{W}_{\left[ N-2\left(
P+1\right) \right] }\mathbf{\delta }_{\left[ 2\left( P+1\right) \right]
}\right\} -\frac{S_{N-2P}}{r^{2}}\left\{ \mathbf{W}_{\left[ N-2P\right] }%
\mathbf{\delta }_{\left[ 2P\right] }\right\}  \label{LzeroWW}
\end{eqnarray}%
Now, determine $R\left( N,P\right) $\ in (\ref{LzeroWW}) using the fact that 
$\left\{ \text{\textbf{\ \ \ }}\right\} $\ does not necessarily commute with
addition. \ Consider the following example: $\ \left\{ W_{\left[ 2\right]
}\delta _{\left[ 2\right] }\right\} $ has $\binom{4}{2}=\allowbreak 6$
terms, i.e., the number of ways to chose 2 indecies from 4 indecies without
repetition. \ Each of the 6 terms has the form $\delta _{ij}\frac{r_{k}}{r}%
\frac{r_{l}}{r}$. \ 

\begin{equation*}
\frac{\partial }{\partial r_{n}}\frac{\partial }{\partial r_{n}}\left(
\delta _{ij}\frac{r_{k}}{r}\frac{r_{l}}{r}\right) =\frac{2}{r^{2}}\delta
_{ij}\left( \delta _{kl}-3\frac{r_{k}}{r}\frac{r_{l}}{r}\right)
\end{equation*}%
Thus,%
\begin{eqnarray*}
\frac{\partial }{\partial r_{n}}\frac{\partial }{\partial r_{n}}\left\{ W_{%
\left[ 2\right] }\delta _{\left[ 2\right] }\right\} &=&\frac{\partial }{%
\partial r_{n}}\frac{\partial }{\partial r_{n}}\left\{ \delta _{ij}\frac{%
r_{k}}{r}\frac{r_{l}}{r}\right\} =\left\{ \frac{\partial }{\partial r_{n}}%
\frac{\partial }{\partial r_{n}}\left( \delta _{ij}\frac{r_{k}}{r}\frac{r_{l}%
}{r}\right) \right\} \\
&=&\frac{2}{r^{2}}\left\{ \delta _{ij}\left( \delta _{kl}-3\frac{r_{k}}{r}%
\frac{r_{l}}{r}\right) \right\}
\end{eqnarray*}%
Now, $\left\{ \delta _{ij}\left( \delta _{kl}-3\frac{r_{k}}{r}\frac{r_{l}}{r}%
\right) \right\} $\ \ is the sum of 6 distinct terms of type $\delta
_{ij}\left( \delta _{kl}-3\frac{r_{k}}{r}\frac{r_{l}}{r}\right) $. \ If we
use of the distributive law of multiplication then we have 6 distinct terms
of type $-3\delta _{ij}\frac{r_{k}}{r}\frac{r_{l}}{r}$,\ and also 6 terms of
type $\delta _{ij}\delta _{kl}$\ , but the latter has repeated terms, only 3
are distinct. \ Thus, 
\begin{equation*}
\frac{2}{r^{2}}\left\{ \delta _{ij}\left( \delta _{kl}-3\frac{r_{k}}{r}\frac{%
r_{l}}{r}\right) \right\} =\frac{2}{r^{2}}\left\{ \delta _{ij}\delta
_{kl}-3\delta _{ij}\frac{r_{k}}{r}\frac{r_{l}}{r}\right\} =\frac{2}{r^{2}}%
\left( 2\left\{ \delta _{ij}\delta _{kl}\right\} -3\left\{ \delta _{ij}\frac{%
r_{k}}{r}\frac{r_{l}}{r}\right\} \right) .
\end{equation*}%
The multiplier $2$\ on term $2\left\{ \delta _{ij}\delta _{kl}\right\} $ is
the number of terms in $\left\{ \delta _{ij}\delta _{kl}-3\delta _{ij}\frac{%
r_{k}}{r}\frac{r_{l}}{r}\right\} $, which is 6, divided by the number of
terms in $\left\{ \delta _{ij}\delta _{kl}\right\} $, which is 3. \ The
number of terms in $\left\{ \delta _{ij}\delta _{kl}-3\delta _{ij}\frac{r_{k}%
}{r}\frac{r_{l}}{r}\right\} $ is the same as the number of terms in $\left\{
\delta _{ij}\frac{r_{k}}{r}\frac{r_{l}}{r}\right\} $. \ Thus, to determine $%
R\left( N,P\right) $\ in (\ref{LzeroWW}), requires the number of terms in $%
\left\{ \mathbf{W}_{\left[ N-2P\right] }\mathbf{\delta }_{\left[ 2P\right]
}\right\} $\ divided by the number of terms in $\left\{ \mathbf{W}_{\left[
N-2\left( P+1\right) \right] }\mathbf{\delta }_{\left[ 2\left( P+1\right) %
\right] }\right\} $. \ From (\ref{numbterms}), $\left\{ \mathbf{W}_{\left[ N%
\right] }\mathbf{\delta }_{\left[ 2P\right] }\right\} $ has $\binom{2P+N}{N}%
\left( 2P-1\right) !!$ terms. \ Thus, the number of terms in $\left\{ 
\mathbf{W}_{\left[ N-2P\right] }\mathbf{\delta }_{\left[ 2P\right] }\right\} 
$ is

\begin{equation*}
\binom{2P+N-2P}{N-2P}\left( 2P-1\right) !!=\binom{N}{N-2P}\left( 2P-1\right)
!!
\end{equation*}%
The number of terms in $\mathbf{W}_{\left[ N-2\left( P+1\right) \right] }%
\mathbf{\delta }_{\left[ 2\left( P+1\right) \right] }$ is%
\begin{eqnarray*}
\binom{2\left( P+1\right) +N-2\left( P+1\right) }{N-2\left( P+1\right) }%
\left( 2\left( P+1\right) -1\right) !! &=&\binom{N}{N-2\left( P+1\right) }%
\left( 2P+1\right) !! \\
&=&\binom{N}{2\left( P+1\right) }\left( 2P+1\right) !!
\end{eqnarray*}%
Thus, the number of terms in $\left\{ \mathbf{W}_{\left[ N-2P\right] }%
\mathbf{\delta }_{\left[ 2P\right] }\right\} $\ divided by the number of
terms in $\mathbf{W}_{\left[ N-2\left( P+1\right) \right] }\mathbf{\delta }_{%
\left[ 2\left( P+1\right) \right] }$\ is%
\begin{equation*}
\frac{\binom{N}{N-2P}\left( 2P-1\right) !!}{\binom{N}{2\left( P+1\right) }%
\left( 2P+1\right) !!}
\end{equation*}%
Comparing (\ref{LzeroV}) with (\ref{LzeroWW}) we require that%
\begin{eqnarray}
R\left( N,P\right) &\equiv &2\binom{N-2P}{2}\frac{\binom{N}{N-2P}\left(
2P-1\right) !!}{\binom{N}{2\left( P+1\right) }\left( 2P+1\right) !!}  \notag
\\
&=&\left( 2P+2\right)  \label{R(N,P) definition}
\end{eqnarray}%
In (\ref{R(N,P) definition}), the definition of the binomial coefficient as
factors of factorials and the definition of the double factorial has
resulted in an amazing simplification of $R\left( N,P\right) $.

\qquad The Laplacian of the product of two functions $f$\ and $g$\ is (\ref%
{1Laplaceident}). \ When applied to (\ref{isoformula}), the case $%
f=A_{P}\left( r\right) $\ and $g=\left\{ \mathbf{W}_{\left[ N-2P\right] }%
\mathbf{\delta }_{\left[ 2P\right] }\right\} $\ is needed. \ Recall that $%
\partial _{r_{i}}A\left( r\right) =\frac{r_{i}}{r}\partial _{r}A\left(
r\right) $. \ The last term in (\ref{1Laplaceident})\ vanishes\ as follows: $%
\left( \partial _{r_{i}}f\right) \left( \partial _{r_{i}}g\right) =\left[ 
\frac{1}{r}\partial _{r}A_{P}\left( r\right) \right] r_{i}\partial
_{r_{i}}\left\{ \mathbf{W}_{\left[ N-2P\right] }\mathbf{\delta }_{\left[ 2P%
\right] }\right\} =\left[ \frac{1}{r}\partial _{r}A\left( r\right) \right]
\left\{ \left[ r_{i}\partial _{r_{i}}\mathbf{W}_{\left[ N-2P\right] }\right] 
\mathbf{\delta }_{\left[ 2P\right] }\right\} =0$; this vanishes because (\ref%
{helps}) shows that $r_{i}\partial _{r_{i}}\mathbf{W}_{\left[ N-2P\right]
}=0 $. \ Then (\ref{LzeroWW}) used in (\ref{1Laplaceident})\ combined with $%
\nabla ^{2}A\left( r\right) =\left( \partial _{r}^{2}+\frac{2}{r}\partial
_{r}\right) A\left( r\right) $\ give

\begin{eqnarray}
\nabla ^{2}\left[ A_{P}\left( r\right) \left\{ \mathbf{W}_{\left[ N-2P\right]
}\mathbf{\delta }_{\left[ 2P\right] }\right\} \right] &=&\left[ \left(
\partial _{r}^{2}+\frac{2}{r}\partial _{r}-\frac{S_{N-2P}}{r^{2}}\right)
A_{P}\left( r\right) \right] \left\{ \mathbf{W}_{\left[ N-2P\right] }\mathbf{%
\delta }_{\left[ 2P\right] }\right\}  \notag \\
&&+A_{P}\left( r\right) \frac{R\left( N,P\right) }{r^{2}}\left\{ \mathbf{W}_{%
\left[ N-2\left( P+1\right) \right] }\mathbf{\delta }_{\left[ 2\left(
P+1\right) \right] }\right\} .  \label{LAPtotal}
\end{eqnarray}
The Laplacian operation on (\ref{isoformula}) is simply the sum, $\overset{M}%
{\underset{P=0}{\sum }}$, of terms (\ref{LAPtotal}).

\section{APPENDIX F: \ Matrix Algorithms}

\qquad For computations, it is useful to write (\ref{1isoformula}) as a
matrix equation. \ Let the column index be $J\equiv P+1$, and the row index
be $I\equiv \left( N_{2}/2\right) +1$, such that both $J$ and $I$ range from 
$1$ to $M+1$ in (\ref{1isoformula}). \ Use $N_{3}=0$ in (\ref{0sinisocoefs}-%
\ref{1isocoefs}) to define the following matrix elements 
\begin{equation}
M_{N}\left( I,J\right) =0\text{, for }J<I\text{, \ i.e., }M_{N}\left(
I,J\right) =0\text{ below the main diagonal;}  \label{0Melements}
\end{equation}%
whereas for $J\geq I,$%
\begin{equation}
M_{N}\left( I,J\right) =\left( N-2I+2\right) !\left( 2I-2\right) !/\left[
\left( N-2\left( J-1\right) \right) !2^{J-1}\left( I-1\right) !\left(
J-I\right) !\right] \text{.}  \label{1Melements1}
\end{equation}%
The chosen $M+1$ linearly independent components of $\mathbf{D}_{\left[ N%
\right] }$\ are arranged in a column vector having $D_{\left[ N:N-2I+2,2I-2,0%
\right] }$\ as in its $I$-th row, and the $M+1$ scalar functions $D_{N,P}$
are likewise arranged in a column vector having $D_{N,I-1}$\ in its $I$-th
row. \ Then (\ref{1isoformula}) is written as the matrix equation 
\begin{equation}
\left( 
\begin{array}{c}
D_{\left[ N:N,0,0\right] } \\ 
D_{\left[ N:N-2,2,0\right] } \\ 
\mathbf{\vdots } \\ 
D_{\left[ N:N-2M,2M,0\right] }%
\end{array}%
\right) =\left( 
\begin{array}{cccc}
M_{N}\left( 1,1\right) & M_{N}\left( 1,2\right) & \cdots & M_{N}\left(
1,M+1\right) \\ 
0 & M_{N}\left( 2,2\right) & \cdots & M_{N}\left( 2,M+1\right) \\ 
\vdots & \vdots & \mathbf{\ddots } & \vdots \\ 
0 & 0 & \cdots & M_{N}\left( M+1,M+1\right)%
\end{array}%
\right) \left( 
\begin{array}{c}
D_{N,0} \\ 
D_{N,1} \\ 
\mathbf{\vdots } \\ 
D_{N,M}%
\end{array}%
\right) .  \label{1Dmatrixeq}
\end{equation}%
Denote a matrix having matrix elements $A\left( I,J\right) $ by \frame{$%
A\left( I,J\right) $}\ \ Then (\ref{1Dmatrixeq}) and its solution are
(respectively) 
\begin{equation}
\frame{$D_{\left[ N:N-2J+2,2J-2,0\right] }$}=\frame{$M_{N}\left( I,J\right) $%
}\text{ }\frame{$D_{N,I-1}$},\text{ and }\frame{$D_{N,I-1}$}=\frame{$%
M_{N}\left( I,J\right) $}^{-1}\frame{$D_{\left[ N:N-2J+2,2J-2,0\right] }$},
\label{1Dsolveq}
\end{equation}%
where $\frame{$M_{N}\left( I,J\right) $}^{-1}$ is the inverse of $\frame{$%
M_{N}\left( I,J\right) $}$. \ The determinant of $\frame{$M_{N}\left(
I,J\right) $}$\ is the product of its diagonal elements; from (\ref%
{1Melements1}) that product is nonzero, hence $\frame{$M_{N}\left(
I,J\right) $}^{-1}$\ exists. \ This inverse matrix is to be calculated
numerically. \ In effect, evaluation of the components $D_{\left[
N:N-2J+2,2J-2,0\right] }$\ by means of experimental data or DNS data and use
of the solution in (\ref{1Dsolveq})\ produces the $D_{N,P}$\ for use in (\ref%
{1isoformula}) to completely specify $\mathbf{D}_{\left[ N\right] }$.

\qquad A matrix algorithm is useful for determining the isotropic formula
for the first-order divergence $\nabla _{\mathbf{r}}\cdot \mathbf{D}_{\left[
N+1\right] }$. \ By replacing $N$ by $N+1$ and the symbol $A$ by $D$ in the
divergence formula (\ref{1divDn}), we have 
\begin{eqnarray}
\nabla _{\mathbf{r}}\cdot \mathbf{D}_{\left[ N+1\right] } &=&\overset{%
M^{\prime }}{\underset{P=0}{\sum }}\left\{ \mathbf{W}_{\left[ N-2\left(
P-1\right) \right] }\mathbf{\delta }_{\left[ 2\left( P-1\right) \right]
}\right\} O_{B}\left( N+1,P\right) D_{N+1,P}  \notag \\
&&+\overset{M^{\prime }}{\underset{P=0}{\sum }}\left\{ \mathbf{W}_{\left[
N-2P\right] }\mathbf{\delta }_{\left[ 2P\right] }\right\} O_{C}\left(
P\right) D_{N+1,P},  \label{1divergence} \\
O_{B}\left( N+1,P\right) &\equiv &\left( \left( N+1\right) -2P+1\right)
\left( \partial _{r}-\frac{\left( N+1\right) -2P}{r}\right) ,  \label{1oper}
\\
O_{C}\left( P\right) &\equiv &\left[ \partial _{r}+\frac{2\left( P+1\right) 
}{r}\right] ,  \label{12ndoper} \\
M^{\prime } &=&N/2\text{ if }N\text{ is even; }M^{\prime }=1+\left(
N-1\right) /2\text{ if }N\text{ is odd.}  \label{1Mprime}
\end{eqnarray}%
The differential operators, i.e., $\partial _{r}\equiv \partial /\partial r$%
, in (\ref{1oper}-\ref{12ndoper}) are obtained from (\ref{1operator}-\ref%
{2operator}), and (\ref{1Mprime}) is obtained by replacing $N$ by $N+1$ in (%
\ref{1M}) and simplifying and rearranging the terms. \ Comparison of (\ref%
{1Mprime}) with (\ref{1M}) shows that if $N$\ is even then $M^{\prime }=M$;
thus the matrix representation of $\left\{ \mathbf{W}_{\left[ N-2P\right] }%
\mathbf{\delta }_{\left[ 2P\right] }\right\} \ $within (\ref{1divergence})
is the same as in (\ref{1Dmatrixeq}), which representation was abbreviated
by $\frame{$M_{N}\left( I,J\right) $}$ above. \ On the other hand, if $N$ is
odd, then $M^{\prime }=M+1$, and the last column of the matrix
representation of $\left\{ \mathbf{W}_{\left[ N-2P\right] }\mathbf{\delta }_{%
\left[ 2P\right] }\right\} \ $within (\ref{1divergence}) corresponds to $%
P=M^{\prime }=1+\left( N-1\right) /2$, in which case $\left\{ \mathbf{W}_{%
\left[ N-2P\right] }\mathbf{\delta }_{\left[ 2P\right] }\right\} $ contains $%
\mathbf{W}_{\left[ -1\right] }=0$ such that the last column of the matrix is
zero. \ Thus, the matrix representation of $\left\{ \mathbf{W}_{\left[ N-2P%
\right] }\mathbf{\delta }_{\left[ 2P\right] }\right\} \ $within (\ref%
{1divergence}) is 
\begin{eqnarray}
\frame{$M_{N}^{\ast }\left( I,J\right) $} &=&\left( 
\begin{array}{ccc}
&  & 0 \\ 
& \frame{$M_{N}\left( I,J\right) $} & \vdots \\ 
&  & 0%
\end{array}%
\right) \ \text{if }N\text{\ is odd;} \\
\frame{$M_{N}^{\ast }\left( I,J\right) $} &=&\frame{$M_{N}\left( I,J\right) $%
}\text{ \ if }N\text{\ is even.}
\end{eqnarray}

\qquad In addition to the coefficient $\left\{ \mathbf{W}_{\left[ N-2P\right]
}\mathbf{\delta }_{\left[ 2P\right] }\right\} $, (\ref{1divergence})
contains the coefficient $\left\{ \mathbf{W}_{\left[ N-2\left( P-1\right) %
\right] }\mathbf{\delta }_{\left[ 2\left( P-1\right) \right] }\right\} $. \
From the matrix representation of $\left\{ \mathbf{W}_{\left[ N-2P\right] }%
\mathbf{\delta }_{\left[ 2P\right] }\right\} $, namely (\ref{0Melements}-\ref%
{1Melements1}), the matrix representation of $\left\{ \mathbf{W}_{\left[
N-2\left( P-1\right) \right] }\mathbf{\delta }_{\left[ 2\left( P-1\right) %
\right] }\right\} $ is (recall that $J\equiv P+1$)

\begin{eqnarray*}
M_{N}^{\prime }\left( I,J\right) &=&0\text{, for }J-1<I\text{, \ i.e., }%
M_{N}^{\prime }\left( I,J\right) =0\text{ on and below the main diagonal;} \\
\text{whereas \ for }J &\geq &I, \\
M_{N}^{\prime }\left( I,J\right) &=&\left( N-2I+2\right) !\left( 2I-2\right)
!/\left[ \left( N-2\left( J-2\right) \right) !2^{J-2}\left( I-1\right)
!\left( J-1-I\right) !\right] \text{.}
\end{eqnarray*}%
The matrix having these elements is denoted by $\frame{$M_{N}^{\prime
}\left( I,J\right) $}$. \ Because of (\ref{1Mprime}), if $N$\ is odd, then
the matrix $\frame{$M_{N}^{\prime }\left( I,J\right) $}$ contains the matrix 
$\frame{$M_{N}\left( I,J\right) $}$ shifted to the right by one column and a
first column of zeros is included; that is, 
\begin{equation}
\frame{$M_{N}^{\prime }\left( I,J\right) $}=\left( 
\begin{array}{ccc}
0 &  &  \\ 
\vdots & \frame{$M_{N}\left( I,J-1\right) $} &  \\ 
0 &  & 
\end{array}%
\right) \text{ \ if }N\text{\ is odd.}
\end{equation}%
Because of (\ref{1Mprime}), the same is true if $N$\ is even except that the
right-most column of $\frame{$M_{N}\left( I,J\right) $}$\ is discarded. \
Thus, 
\begin{equation}
\frame{$M_{N}^{\prime }\left( I,J\right) $}=\left( 
\begin{array}{cc}
0 &  \\ 
\vdots & \frame{$M_{N}\left( I,J-1\right) $} \\ 
0 & 
\end{array}%
\right) \text{ \ if }N\text{\ is even (discard the right-most column).}
\end{equation}

\qquad Define operator matrices that are of dimension $M^{\prime }+1$\ by $%
M^{\prime }+1$, that have zeros off of the diagonal, and that have the
operators (\ref{1oper}-\ref{12ndoper}) on the diagonals. \ Thus, recall that 
$J\equiv P+1$, and that $\partial _{r}\equiv \partial /\partial _{r}$, and
define matrix elements 
\begin{equation}
B\left( I,J\right) \equiv \delta _{IJ}\left( N-2J+4\right) \left( \partial
_{r}-\frac{N-2J+3}{r}\right) \text{ and }C\left( I,J\right) \equiv \delta
_{IJ}\left( \partial _{r}+\frac{2J}{r}\right) .  \label{1BCdef}
\end{equation}
The matrices corresponding to $O_{B}\left( N+1,P\right) $\ and $O_{C}\left(
P\right) $ in (\ref{1oper}-\ref{12ndoper}) are denoted by \frame{$B\left(
I,J\right) $}, and $\frame{$C\left( I,J\right) $}$, respectively.

\qquad Let the components of $\nabla _{\mathbf{r}}\cdot \mathbf{D}_{\left[
N+1\right] }$\ be denoted by $\left( \nabla _{\mathbf{r}}\cdot \mathbf{D}_{%
\left[ N+1\right] }\right) _{\left[ N:N_{1},N_{2},N_{3}\right] }$, which
denotes the fact that $\nabla _{\mathbf{r}}\cdot \mathbf{D}_{\left[ N+1%
\right] }$\ is a tensor of order $N$. \ In matrix notation, (\ref%
{1divergence}) gives 
\begin{equation}
\left( 
\begin{array}{c}
\left( \nabla _{\mathbf{r}}\cdot \mathbf{D}_{\left[ N+1\right] }\right) _{%
\left[ N:N,0,0\right] } \\ 
\left( \nabla _{\mathbf{r}}\cdot \mathbf{D}_{\left[ N+1\right] }\right) _{%
\left[ N:N-2,2,0\right] } \\ 
\mathbf{\vdots } \\ 
\left( \nabla _{\mathbf{r}}\cdot \mathbf{D}_{\left[ N+1\right] }\right) _{%
\left[ N:N-2M,2M,0\right] }%
\end{array}
\right) =\left[ \frame{$M_{N}^{\prime }\left( I,J\right) $}\text{ }\frame{$%
B\left( I,J\right) $}+\frame{$M_{N}^{\ast }\left( I,J\right) $}\text{ }%
\frame{$C\left( I,J\right) $}\right] \left( 
\begin{array}{c}
D_{N+1,0} \\ 
D_{N+1,1} \\ 
\mathbf{\vdots } \\ 
D_{N+1,M^{\prime }}%
\end{array}
\right) .  \label{1divmatr}
\end{equation}

When applied to $\mathbf{D}_{\left[ N+1\right] }$, the solution of (\ref%
{1Dmatrixeq}) is 
\begin{equation*}
\frame{$D_{N+1,I-1}$}=\frame{$M_{N+1}\left( I,J\right) $}^{-1}\frame{$D_{%
\left[ N+1:N+1-2J+2,2J-2,0\right] }$},
\end{equation*}%
substitution of which into (\ref{1divmatr}) gives 
\begin{equation}
\left( 
\begin{array}{c}
\left( \nabla _{\mathbf{r}}\cdot \mathbf{D}_{\left[ N+1\right] }\right) _{%
\left[ N:N,0,0\right] } \\ 
\left( \nabla _{\mathbf{r}}\cdot \mathbf{D}_{\left[ N+1\right] }\right) _{%
\left[ N:N-2,2,0\right] } \\ 
\mathbf{\vdots } \\ 
\left( \nabla _{\mathbf{r}}\cdot \mathbf{D}_{\left[ N+1\right] }\right) _{%
\left[ N:N-2M,2M,0\right] }%
\end{array}%
\right) =\frame{$Y\left( I,J\right) $}\left( 
\begin{array}{c}
D_{\left[ N+1:N+1,0,0\right] } \\ 
D_{\left[ N+1:N+1-2,2,0\right] } \\ 
\mathbf{\vdots } \\ 
D_{\left[ N+1:N+1-2M^{\prime },2M^{\prime },0\right] }%
\end{array}%
\right) ,  \label{1Ymatmult}
\end{equation}%
\begin{equation*}
\text{where, }
\end{equation*}%
\begin{equation}
\frame{$Y\left( I,J\right) $}\equiv \left[ \frame{$M_{N}^{\prime }\left(
I,J\right) $}\text{ }\frame{$B\left( I,J\right) $}+\frame{$M_{N}^{\ast
}\left( I,J\right) $}\text{ }\frame{$C\left( I,J\right) $}\right] \frame{$%
M_{N+1}\left( I,J\right) $}^{-1}.  \label{1Ymatrix2}
\end{equation}%
We see that $\frame{$Y\left( I,J\right) $}$\ is the operator matrix that
operates on the column matrix representation of $\mathbf{D}_{\left[ N+1%
\right] }$\ to produce $\nabla _{\mathbf{r}}\cdot \mathbf{D}_{\left[ N+1%
\right] }$; this is true for any completely symmetric isotropic tensor, not
just true for $\mathbf{D}_{\left[ N+1\right] }$.

\qquad It is helpful to illustrate this algorithm for $N=2$ and $N=3$. \ Two
examples are needed because the algorithm differs for even $N$ as compared
to odd $N$. \ For $N=2$, (\ref{1Ymatrix2}) is%
\begin{equation*}
\frame{$Y\left( I,J\right) $}=
\end{equation*}

\begin{eqnarray}
&&\left[ 
\begin{array}{c}
\left( 
\begin{array}{cc}
0 & M_{2}\left( 1,1\right) \\ 
0 & 0%
\end{array}%
\right) \left( 
\begin{array}{cc}
B\left( 1,1\right) & 0 \\ 
0 & B\left( 2,2\right)%
\end{array}%
\right) + \\ 
+\left( 
\begin{array}{cc}
M_{2}\left( 1,1\right) & M_{2}\left( 1,2\right) \\ 
0 & M_{2}\left( 2,2\right)%
\end{array}%
\right) \left( 
\begin{array}{cc}
C\left( 1,1\right) & 0 \\ 
0 & C\left( 2,2\right)%
\end{array}%
\right)%
\end{array}%
\right] \cdot \left( 
\begin{array}{cc}
M_{3}\left( 1,1\right) & M_{3}\left( 1,2\right) \\ 
0 & M_{3}\left( 2,2\right)%
\end{array}%
\right) ^{-1}  \label{1Y2} \\
&=&\left( 
\begin{array}{cc}
\partial _{r}+\frac{2}{r} & -\frac{4}{r} \\ 
0 & \partial _{r}+\frac{4}{r}%
\end{array}%
\right) .  \label{1Y2nd}
\end{eqnarray}%
Computer evaluation of (\ref{1Y2}) produced (\ref{1Y2nd}). \ Consequently, (%
\ref{1Ymatmult}) is 
\begin{equation}
\left( 
\begin{array}{c}
\left( \nabla _{\mathbf{r}}\cdot \mathbf{D}_{\left[ 3\right] }\right) _{%
\left[ 2:2,0,0\right] } \\ 
\left( \nabla _{\mathbf{r}}\cdot \mathbf{D}_{\left[ 3\right] }\right) _{%
\left[ 2:0,2,0\right] }%
\end{array}%
\right) =\left( 
\begin{array}{cc}
\partial _{r}+\frac{2}{r} & -\frac{4}{r} \\ 
0 & \partial _{r}+\frac{4}{r}%
\end{array}%
\right) \left( 
\begin{array}{c}
D_{\left[ 3:3,0,0\right] } \\ 
D_{\left[ 3:1,2,0\right] }%
\end{array}%
\right) =\left( 
\begin{array}{c}
\left( \partial _{r}+\frac{2}{r}\right) D_{111}-\frac{4}{r}D_{122} \\ 
\left( \partial _{r}+\frac{4}{r}\right) D_{122}%
\end{array}%
\right) ,
\end{equation}%
where explicit-index notation is given at far right by use of $D_{\left[
3:3,0,0\right] }\equiv D_{111}$ and $D_{\left[ 3:1,2,0\right] }\equiv
D_{122} $.

\qquad For $N=3$, $\frame{$Y\left( I,J\right) $}$ from (\ref{1Ymatrix2}) is 
\begin{eqnarray*}
&&\left[ 
\begin{array}{c}
\left( 
\begin{array}{ccc}
0 & M_{3}\left( 1,1\right) & M_{3}\left( 1,2\right) \\ 
0 & 0 & M_{3}\left( 2,2\right)%
\end{array}%
\right) \left( 
\begin{array}{ccc}
B\left( 1,1\right) & 0 & 0 \\ 
0 & B\left( 2,2\right) & 0 \\ 
0 & 0 & B\left( 3,3\right)%
\end{array}%
\right) + \\ 
\left( 
\begin{array}{ccc}
M_{3}\left( 1,1\right) & M_{3}\left( 1,2\right) & 0 \\ 
0 & M_{3}\left( 2,2\right) & 0%
\end{array}%
\right) \left( 
\begin{array}{ccc}
C\left( 1,1\right) & 0 & 0 \\ 
0 & C\left( 2,2\right) & 0 \\ 
0 & 0 & C\left( 3,3\right)%
\end{array}%
\right)%
\end{array}%
\right] \cdot \\
&&\cdot \left( 
\begin{array}{ccc}
M_{4}\left( 1,1\right) & M_{4}\left( 1,2\right) & M_{4}\left( 1,3\right) \\ 
0 & M_{4}\left( 2,2\right) & M_{4}\left( 2,3\right) \\ 
0 & 0 & M_{4}\left( 3,3\right)%
\end{array}%
\right) ^{-1} \\
&=&\left( 
\begin{array}{ccc}
\partial _{r}+\frac{2}{r} & -\frac{6}{r} & 0 \\ 
0 & \partial _{r}+\frac{4}{r} & -\frac{4}{3r}%
\end{array}%
\right)
\end{eqnarray*}%
As with (\ref{1Y2nd}), the matrix was evaluated using a computer program. \
Consequently, (\ref{1Ymatmult}) is 
\begin{eqnarray}
\left( 
\begin{array}{c}
\left( \nabla _{\mathbf{r}}\cdot \mathbf{D}_{\left[ 4\right] }\right) _{%
\left[ 3:3,0,0\right] } \\ 
\left( \nabla _{\mathbf{r}}\cdot \mathbf{D}_{\left[ 4\right] }\right) _{%
\left[ 3:1,2,0\right] }%
\end{array}%
\right) &=&\left( 
\begin{array}{ccc}
\partial _{r}+\frac{2}{r} & -\frac{6}{r} & 0 \\ 
0 & \partial _{r}+\frac{4}{r} & -\frac{4}{3r}%
\end{array}%
\right) \left( 
\begin{array}{c}
D_{\left[ 4:4,0,0\right] } \\ 
D_{\left[ 4:2,2,0\right] } \\ 
D_{\left[ 4:0,4,0\right] }%
\end{array}%
\right)  \notag \\
&=&\left( 
\begin{array}{c}
\left( \partial _{r}+\frac{2}{r}\right) D_{1111}-\frac{6}{r}D_{1122} \\ 
\left( \partial _{r}+\frac{4}{r}\right) D_{1122}-\frac{4}{3r}D_{2222}%
\end{array}%
\right) ,  \label{13rdYeval}
\end{eqnarray}%
where explicit-index notation is used in (\ref{13rdYeval}).

\qquad A matrix algorithm is also needed for the Laplacian of a symmetric
tensor. \ Performing the Laplacian of (\ref{isoformula}) and use of (\ref%
{LAPtotal}) gives 
\begin{equation}
\nabla _{\mathbf{r}}^{2}\mathbf{D}_{\left[ N\right] }\left( \mathbf{r}%
\right) =\overset{M}{\underset{P=0}{\sum }}\left( 
\begin{array}{c}
\left\{ \mathbf{W}_{\left[ N-2P\right] }\mathbf{\delta }_{\left[ 2P\right]
}\right\} \left( \partial _{r}^{2}+\frac{2}{r}\partial _{r}-\frac{S_{N-2P}}{%
r^{2}}\right) D_{N,P} \\ 
+\left\{ \mathbf{W}_{\left[ N-2\left( P+1\right) \right] }\delta _{\left[
2\left( P+1\right) \right] }\right\} \frac{R\left( N,P\right) }{r^{2}}D_{N,P}%
\end{array}%
\right) .  \label{Lap1D}
\end{equation}%
It is necessary to recall the definitions (\ref{Sdefinition}) and \ref%
{R(N,P) definition}). \ The matrix representation of $\left\{ \mathbf{W}_{%
\left[ N-2P\right] }\mathbf{\delta }_{\left[ 2P\right] }\right\} \ $within (%
\ref{Lap1D}) is the same as in (\ref{0Melements}-\ref{1Melements1}), namely $%
\frame{$M_{N}\left( I,J\right) $}$. \ The matrix representation of $\left\{ 
\mathbf{W}_{\left[ N-2\left( P+1\right) \right] }\delta _{\left[ 2\left(
P+1\right) \right] }\right\} $ within (\ref{Lap1D}) is obtained from (\ref%
{0Melements}-\ref{1Melements1})\ by replacing $J$\ by $J+1$, i.e., 
\begin{equation*}
M_{N}^{\#}\left( I,J\right) =0\text{, for }J+1<I,
\end{equation*}%
whereas\ for $J+1\geq I,$%
\begin{equation}
M_{N}^{\#}\left( I,J\right) =\left( N-2I+2\right) !\left( 2I-2\right) !/ 
\left[ \left( N-2J\right) !2^{J}\left( I-1\right) !\left( J+1-I\right) !%
\right] .  \label{Mhatmatrixele}
\end{equation}%
This is just the square matrix that appears in (\ref{1Dmatrixeq}) except
that the left-most column in (\ref{1Dmatrixeq}) is discarded and the matrix
is then shifted leftward by one column and the right-most column is zeros. \
Those zeros appear because in the right-most column $J=P+1=M+1$ such that $%
M_{N}^{\#}\left( I,M+1\right) $ contains the factor $1/\left( N-2\left(
M+1\right) \right) !$ which is $1/\left( -2\right) !=0$ if $N$ is even and
is $1/\left( -1\right) !=0$ if $N$\ is odd(see Abramowitz and Stegun, 1964,
equation 6.1.7). \ Thus, 
\begin{equation*}
\fbox{$M_{N}^{\#}\left( I,J\right) $}=\left( 
\begin{array}{cc}
& 0 \\ 
M_{N}\left( I,J+1\right) & \vdots \\ 
& 0%
\end{array}%
\right) \text{ (discard the left-most column).}
\end{equation*}%
Define two operator matrices that are zero off the main diagonal and contain 
$\left( \partial _{r}^{2}+\frac{2}{r}\partial _{r}-\frac{S_{N-2P}}{r^{2}}%
\right) $\ and $\frac{R\left( N,P\right) }{r^{2}}$\ on the main diagonal;
i.e., their matrix elements are$\ E\left( I,J\right) =\delta _{IJ}\left(
\partial _{r}^{2}+\frac{2}{r}\partial _{r}-\frac{S_{N-2\left( J-1\right) }}{%
r^{2}}\right) $ and $F\left( I,J\right) =\delta _{IJ}R\left( N,J-1\right)
/r^{2}$. \ Analogous to the derivation of (\ref{1Ymatmult}), the matrix
representation of (\ref{Lap1D}) is 
\begin{eqnarray}
\left( 
\begin{array}{c}
\left( \nabla _{\mathbf{r}}^{2}\mathbf{D}_{\left[ N\right] }\right) _{\left[
N:N,0,0\right] } \\ 
\left( \nabla _{\mathbf{r}}^{2}\mathbf{D}_{\left[ N\right] }\right) _{\left[
N:N-2,2,0\right] } \\ 
\mathbf{\vdots } \\ 
\left( \nabla _{\mathbf{r}}^{2}\mathbf{D}_{\left[ N\right] }\right) _{\left[
N:N-2M,2M,0\right] }%
\end{array}%
\right) &=&\frame{$X\left( I,J\right) $}\left( 
\begin{array}{c}
D_{\left[ N:N,0,0\right] } \\ 
D_{\left[ N:N-2,2,0\right] } \\ 
\mathbf{\vdots } \\ 
D_{\left[ N:N-2M,2M,0\right] }%
\end{array}%
\right) ,  \label{Lapmatrixeq} \\
\text{where, }\frame{$X\left( I,J\right) $} &\equiv &\left[ \frame{$%
M_{N}\left( I,J\right) $}\text{ }\frame{$E\left( I,J\right) $}+\frame{$%
M_{N}^{\#}\left( I,J\right) $}\text{ }\frame{$F\left( I,J\right) $}\right] 
\frame{$M_{N}\left( I,J\right) $}^{-1}.  \notag
\end{eqnarray}%
For both $N=2$ and $N=3$, the matrix representation of $\frame{$X\left(
I,J\right) $}$ is

$\frame{$X\left( I,J\right) $}\equiv \left[ 
\begin{array}{c}
\left( 
\begin{array}{cc}
M_{N}\left( 1,1\right) & M_{N}\left( 1,2\right) \\ 
0 & M_{N}\left( 2,2\right)%
\end{array}
\right) \left( 
\begin{array}{cc}
E\left( 1,1\right) & 0 \\ 
0 & E\left( 2,2\right)%
\end{array}
\right) \\ 
+\left( 
\begin{array}{cc}
M_{N}\left( 1,2\right) & 0 \\ 
M_{N}\left( 2,2\right) & 0%
\end{array}
\right) \left( 
\begin{array}{cc}
F\left( 1,1\right) & 0 \\ 
0 & F\left( 2,2\right)%
\end{array}
\right)%
\end{array}
\right] \left( 
\begin{array}{cc}
M_{N}\left( 1,1\right) & M_{N}\left( 1,2\right) \\ 
0 & M_{N}\left( 2,2\right)%
\end{array}
\right) ^{-1}$

For $N=2$ (\ref{Lapmatrixeq}) is 
\begin{equation*}
\left( 
\begin{array}{c}
\left( \nabla _{\mathbf{r}}^{2}\mathbf{D}_{\left[ 2\right] }\right) _{\left[
2:2,0,0\right] } \\ 
\left( \nabla _{\mathbf{r}}^{2}\mathbf{D}_{\left[ 2\right] }\right) _{\left[
2:0,2,0\right] }%
\end{array}%
\right) =\frame{$X\left( I,J\right) $}\left( 
\begin{array}{c}
D_{\left[ 2:2,0,0\right] } \\ 
D_{\left[ 2:0,2,0\right] }%
\end{array}%
\right) =\left( 
\begin{array}{c}
\left( \nabla _{\mathbf{r}}^{2}\mathbf{D}_{\left[ 2\right] }\right) _{11} \\ 
\left( \nabla _{\mathbf{r}}^{2}\mathbf{D}_{\left[ 2\right] }\right) _{22}%
\end{array}%
\right) =
\end{equation*}%
\begin{equation}
\left( 
\begin{array}{cc}
\partial _{r}^{2}+\frac{2}{r}\partial _{r}-\frac{4}{r^{2}} & \frac{4}{r^{2}}
\\ 
\frac{2}{r^{2}} & \partial _{r}^{2}+\frac{2}{r}\partial _{r}-\frac{2}{r^{2}}%
\end{array}%
\right) \left( 
\begin{array}{c}
D_{11} \\ 
D_{22}%
\end{array}%
\right) =\left( 
\begin{array}{c}
\left( \partial _{r}^{2}+\frac{2}{r}\partial _{r}-\frac{4}{r^{2}}\right)
D_{11}+\frac{4}{r^{2}}D_{22} \\ 
\frac{2}{r^{2}}D_{11}+\left( \partial _{r}^{2}+\frac{2}{r}\partial _{r}-%
\frac{2}{r^{2}}\right) D_{22}%
\end{array}%
\right) ,  \label{LapNis2}
\end{equation}%
where the matrix was evaluated using a computer program. \ For $N=3$ the
matrix algorithm is 
\begin{eqnarray}
\left( 
\begin{array}{c}
\left( \nabla _{\mathbf{r}}^{2}\mathbf{D}_{\left[ 3\right] }\right) _{111}
\\ 
\left( \nabla _{\mathbf{r}}^{2}\mathbf{D}_{\left[ 3\right] }\right) _{122}%
\end{array}%
\right) &=&\frame{$X\left( I,J\right) $}\left( 
\begin{array}{c}
D_{111} \\ 
D_{122}%
\end{array}%
\right) =\left( 
\begin{array}{cc}
\partial _{r}^{2}+\frac{2}{r}\partial _{r}-\frac{6}{r^{2}} & \frac{12}{r^{2}}
\\ 
\frac{2}{r^{2}} & -\frac{8}{r^{2}}+\partial _{r}^{2}+\frac{2}{r}\partial _{r}%
\end{array}%
\right) \left( 
\begin{array}{c}
D_{111} \\ 
D_{122}%
\end{array}%
\right)  \notag \\
&=&\left( 
\begin{array}{c}
\left( \partial _{r}^{2}+\frac{2}{r}\partial _{r}-\frac{6}{r^{2}}\right)
D_{111}+\frac{12}{r^{2}}D_{122} \\ 
\frac{2}{r^{2}}D_{111}+\left( -\frac{8}{r^{2}}+\partial _{r}^{2}+\frac{2}{r}%
\partial _{r}\right) D_{122}%
\end{array}%
\right) .  \label{LapNis3}
\end{eqnarray}

\end{document}